\documentclass[useAMS,usenatbib,letterpaper,10pt]{mn2e}
\usepackage{times}
\usepackage{epsfig}
\usepackage{natbib}
\usepackage{amssymb}
\usepackage{amsmath}
\usepackage{verbatim}
\usepackage[bookmarks,bookmarksnumbered,colorlinks=true, citecolor=blue, linkcolor=black]{hyperref}
\usepackage[dvipsnames]{xcolor}
\usepackage[normalem]{ulem}

\newcommand{\Msun}{M$_\odot$} 
\newcommand{\vphi}{$\bar{v}_\phi$~}

\title[Dynamical Masses from Gas Kinematics]{Measuring dynamical masses from gas kinematics in simulated high-redshift galaxies
}
\author[Wellons et al]{Sarah Wellons$^{1}$\thanks{E-mail: sarah.wellons@northwestern.edu}, Claude-Andr\'e Faucher-Gigu\`ere$^{1}$, Daniel Angl\'es-Alc\'azar$^{2,3}$, \newauthor
Christopher C. Hayward$^{2}$, Robert Feldmann$^{4}$, Philip F. Hopkins$^{5}$, and Du\v{s}an Kere\v{s}$^{6}$ \\
$^{1}$CIERA and Department of Physics and Astronomy, Northwestern University, 2145 Sheridan Road, Evanston, IL 60208, USA \\
$^{2}$Center for Computational Astrophysics, Flatiron Institute, 162 Fifth Avenue, New York, NY 10010, USA \\
$^{3}$Department of Physics, University of Connecticut, 196 Auditorium Road, U-3046, Storrs, CT 06269-3046, USA \\
$^{4}$Institute for Computational Science, University of Zurich, Zurich CH-8057, Switzerland \\
$^{5}$TAPIR, MC 350-17, California Institute of Technology, Pasadena, CA 91125, USA \\
$^{6}$Department of Physics, Center for Astrophysics and Space Sciences, University of California, San Diego, La Jolla, CA, USA}
\begin{document}

\maketitle

\label{firstpage}

\begin{abstract}
Advances in instrumentation have recently extended detailed measurements of gas kinematics to large samples of high-redshift galaxies. 
Relative to most nearby, thin disk galaxies, in which gas rotation accurately traces the gravitational potential, the interstellar medium (ISM) of $z\gtrsim1$ galaxies is typically more dynamic and exhibits elevated turbulence. 
If not properly modeled, these effects can strongly bias dynamical mass measurements.
We use high-resolution FIRE-2 cosmological zoom-in simulations to analyze the physical effects that must be considered to correctly infer dynamical masses from gas kinematics. 
Our analysis covers a range of galaxy properties from low-redshift Milky-Way-mass galaxies to massive high-redshift galaxies ($M_{\star}>10^{11}$ M$_{\odot}$ at $z=1$).  
Selecting only snapshots where a disk is present, we calculate the rotational profile $\bar{v}_\phi(r)$ of the cool ($10^{3.5}~\rm{K}<T<10^{4.5}~\rm{K}$) gas and compare it to the circular velocity $v_c=\sqrt{GM_{\rm enc}/r}$.  In the simulated galaxies, the gas rotation traces the circular velocity at intermediate radii, but the two quantities diverge significantly in the center and in the outer disk.  Our simulations appear to over-predict observed rotational velocities in the centers of massive galaxies (likely from a lack of black hole feedback), so we focus on larger radii.  Gradients in the turbulent pressure at these radii can provide additional radial support and bias dynamical mass measurements low by up to 40\%.
In both the interior and exterior, the gas' motion can be significantly non-circular due to e.g. bars, satellites, and inflows/outflows.  
We discuss the accuracy of commonly-used analytic models for pressure gradients (or ``asymmetric drift") in the ISM of high-redshift galaxies.

\end{abstract}


\section{Introduction}

Measuring the kinematic motion in galaxies and galaxy clusters has long been a valuable means of estimating its mass content.  Famously, \citet{Rubin1978} demonstrated the presence of dark matter on galactic scales by showing that the rotation curve of the Andromeda galaxy is flatter than would be predicted from its baryonic mass distribution alone.  In an earlier well-known result, \citet{Zwicky1933} used the motions of galaxies themselves to infer the mass of the dark matter halos around the Coma cluster.

Similar techniques have been applied to galaxies at a variety of masses and redshifts.
In the last few years, spatially-resolved studies of gas kinematics at high redshift have become commonplace thanks to newly available observational facilities and instruments. 
For example, optical/infrared integral field units (IFUs) such as SINFONI \citep[][]{Eisenhauer2003}, KMOS \citep[][]{Sharples2013}, MUSE \citep[][]{Bacon2010}, OSIRIS \citep[][]{Larkin2006}, and KCWI \citep[][]{Morrissey2018} have been installed on 8-10m telescopes. 
These IFU instruments have already provided crucial new windows into the kinematics of the ionized interstellar medium (ISM) of distant galaxies, including dynamical mass estimates \citep[e.g.,][]{Wuyts2016}, velocity dispersion measurements \citep[e.g.,][]{Law2009, Johnson2018}, and constraints on the properties of outflows from star-forming clumps \citep[e.g.,][]{Newman2012}. 
Although they do not provide as much spatial information, multi-object spectrographs, notably MOSFIRE on Keck \citep[][]{McLean2012}, have also recently enabled order-of-magnitude more efficient kinematic measurements of large samples of high-redshift galaxies \citep[e.g.,][]{Belli2017, Price2019}. 
In the next decade, the quantity and quality of kinematic data on high-redshift galaxies are poised to continue to improve with the advent of 30m-class telescopes. 
Spatially-resolved kinematic studies of the cold ISM of high-redshift galaxies have similarly been recently revolutionized by the high angular resolution and sensitivity of the Atacama Large Millimeter/submillimeter Array (ALMA).

Massive galaxies at high redshift represent an especially intriguing target for kinematic measurements.
These galaxies differ from the local population in several ways, notably in their more compact stellar distributions \citep{Trujillo2007, VanDokkum2008, VanderWel2014}.
This compactness leads to high rotational velocities of 250-350 km/s \citep{VanDokkum2009, Newman2015, Talia2018, Newman2018}, and in some cases even up to 500 km/s \citep[e.g.,][]{VanDokkum2015}.

Accurately inferring dynamical masses from observed rotational velocities is essential for understanding the evolution of galaxies.
For example, \citet{Genzel2017} measured the gas kinematics in galaxies with stellar masses ranging from $0.4 - 1.2 \times 10^{11}$ \Msun~at redshifts from $z = 0.8-2.4$.  They observed velocity profiles which rise and fall with increasing radius, in contrast with the flat or increasing rotation curves common in local spiral galaxies and expected for cold dark matter halos \citep[][]{Sofue2001}. Accounting for turbulent pressure support in the outer disk using a simple analytic model, \citet{Genzel2017} inferred low dark matter fractions of 0-20\% within the galaxies' half-light radii.  Several other studies utilizing gas or stellar kinematics in high-redshift galaxies have also measured dynamical masses which are close to their baryonic masses \citep{Burkert2016, Belli2017, Lang2017, Ubler2018, Price2019}, although see \citet{Drew2018} for a counter-example of a galaxy with a flat rotation curve and dark matter fraction of 0.44 at $z=1.5$. 

These observations suggest a different role for dark matter in high-redshift galaxies relative to the local Universe, but important difficulties remain with the interpretation of these kinematic measurements. 
For example, taken at face value, many of the (total) dynamical masses of $z\sim2$ star-forming galaxies inferred by \cite{Price2019} based on ionized gas velocity dispersions are \emph{smaller} than the estimated mass in baryons alone.
This clearly implies significant errors in at least some of the dynamical mass and/or baryonic mass estimates. 
Even state-of-the-art IFU measurements using adaptive optics are not immune from systematic effects for dynamical mass measurements. 
As \citet{Genzel2017} note, declining rotation curves in the outer parts of galaxies can be due to pressure gradients in the ISM, which induce forces in the opposite direction from gravity. 
Firming up the observational case for strongly baryon-dominated galaxies at high redshift thus requires a solid understanding of how to correct for pressure gradients.

\begin{figure*}
  \centering
  \includegraphics[width=2.1\columnwidth]{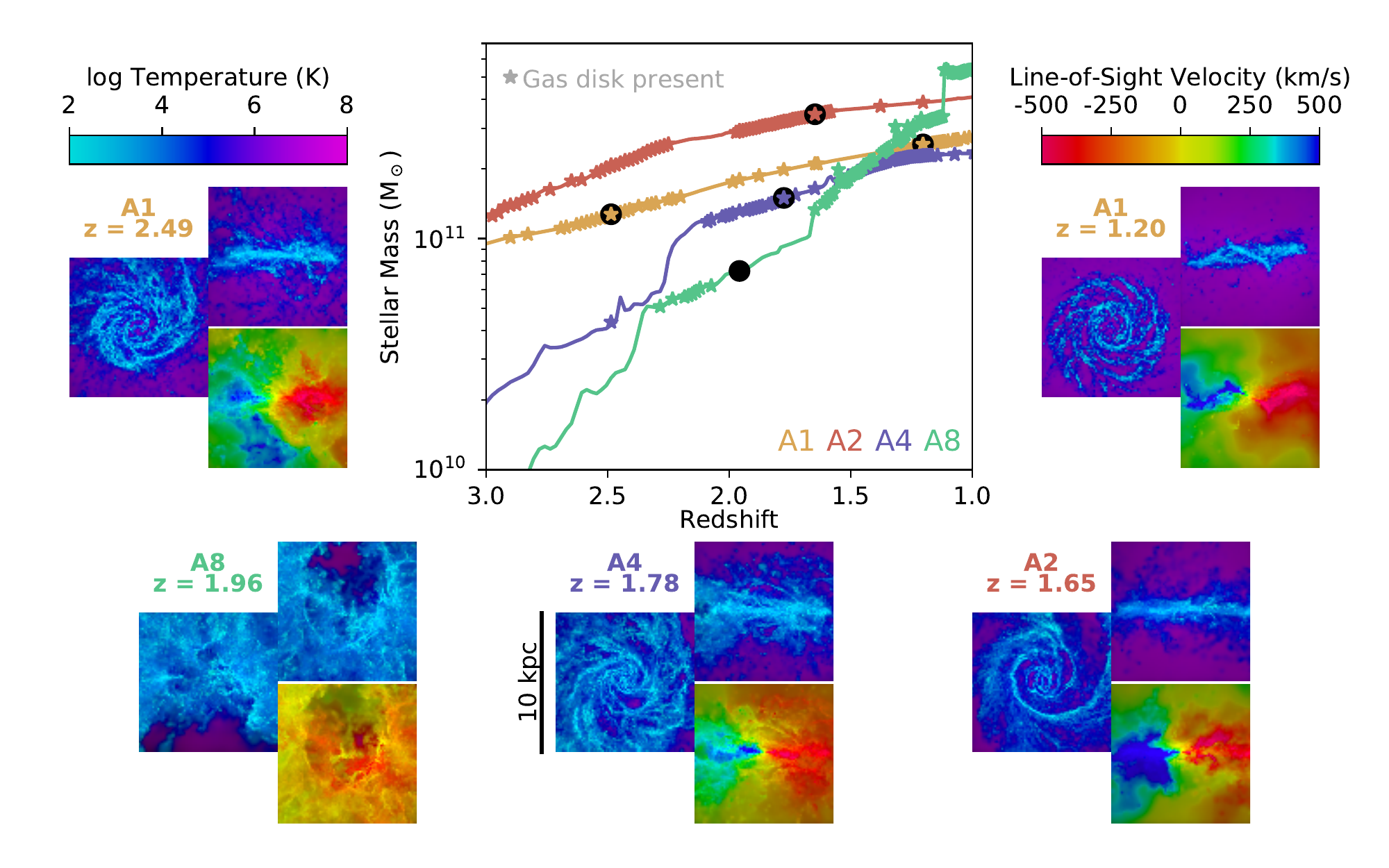}
  \caption{Stellar mass evolution of the central galaxies in the four massive zoom-in simulations from $z = 3$ to $z = 1$.  Stars represent snapshots where a gaseous disk is present, satisfying the criteria described in Section~\ref{ssec:calcs}.  Selected snapshots (designated with black circles) are shown in the surrounding images, all of which have a 10 kpc field of view.  Each cluster of images depicts the gas from a face-on and edge-on perspective, with the hue indicating the gas temperature (mass-weighted along the line of sight) and brightness the gas surface density.  The mass-weighted velocity along the line of sight is also depicted for the edge-on view.  All of the selected snapshots satisfy the ``disk" criteria except the lower left, which shows a snapshot following a galactic outburst.}
  \label{fig:massevol}
\end{figure*}

In this paper, we use cosmological zoom-in simulations to investigate different physical effects important for inferring dynamical masses from measurements of gas rotation curves.
There are many factors, both physical and observational, which can complicate the interpretation of gas rotation curves (e.g. \cite{Pineda2017}). 
On the observational side, a galaxy's inclination and the effects of beam smearing must be accurately accounted for to recover a galaxy's intrinsic rotation from the observed line-of-sight velocity (see e.g. \citet{Wuyts2016} and \citet{Schreiber2018}). 
Various physical effects, such as the pressure gradient mentioned above, can also modify a galaxy's intrinsic rotation such that it is not purely a function of mass content. 

If the ISM is highly turbulent, that turbulence can introduce an additional source of support in the galactic disk via a radial gradient in turbulent pressure.  This effect (sometimes referred to ``asymmetric drift," in an analogy to a similar effect in stellar dynamics) can usually be neglected when the system is highly rotation-dominated, but the messy, turbulent conditions of the high-redshift Universe can make it very important \citep[e.g.,][]{Burkert2010}.  One example of a present-day analog in which the turbulent pressure gradient does afford a significant amount of support is in dwarf galaxies, where the rotational velocity does not greatly exceed the velocity dispersion \citep[see e.g.][]{Valenzuela2007, Read2016}.  In these cases, turbulent pressure can comprise a large fraction of the dynamical support and significantly suppress rotational velocities of the gas.

An additional complication to the inference of mass from rotational velocity arises from assumptions about the shape of the gravitational potential.  The simplest assumption is that of a spherically-symmetric distribution of matter such that rotational velocity is straightforwardly related to the enclosed mass as $\bar{v}_\phi = \sqrt{GM_{\rm enc}/r}$.  In reality, galactic potentials are almost always more complicated.  More generally, the mass distribution may be composed of an ellipsoidal stellar bulge, a disk of gas and stars, and a halo of dark matter and gas which is roughly spherical {but can be significantly triaxial and includes substructure}.  The presence of the disk component modifies the gravitational potential such that the straightforward equation mentioned above cannot be applied.  Observationally, this may be accounted for by fitting a bulge-disk decomposition and combining the gravitational potentials produced by these two components.

A final property of the orbits that must be determined to infer mass from velocity is eccentricity.  In the main body of the disk, we expect the assumption that the gas is moving on circular orbits to hold, since the gas will circularize through self-interaction.  However, when this assumption does not hold, it can lead to misleading results since the orbital velocities of gas on eccentric orbits can differ substantially from the circular velocity.  The assumption of circularity is most likely to break down in the inner region of the disk where non-axisymmetric structures such as bars can be important, and on the outskirts where the inflowing gas has not yet circularized.

\begin{figure*}
  \centering
  \includegraphics[width=1.9\columnwidth]{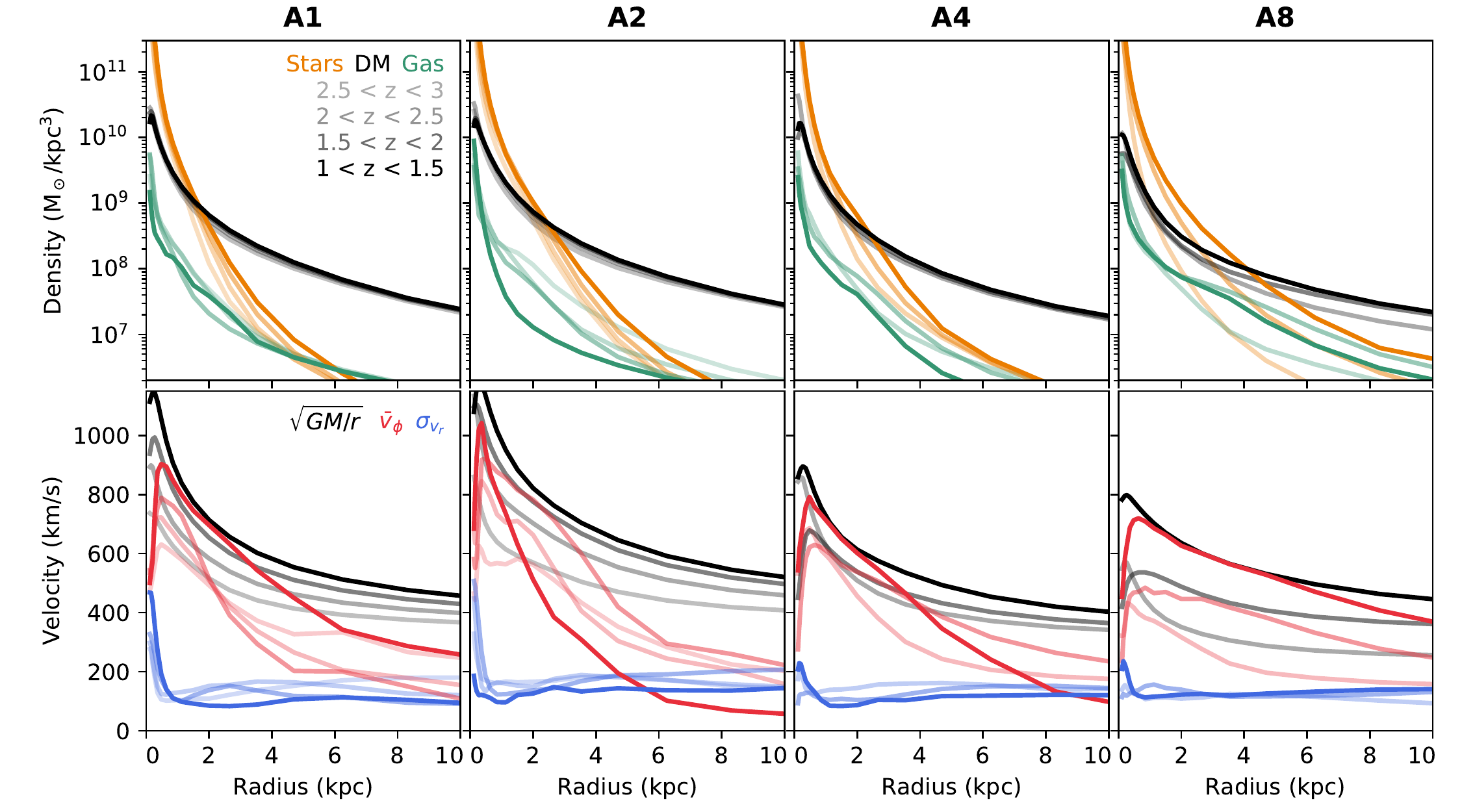}
  \caption{\textbf{Top:} Density profiles of the stellar (orange), dark matter (black), and gas (teal) content of the simulated galaxies.  Each component is shown up to four times, showing averages in the redshift intervals between $z$ = 3, 2.5, 2, 1.5, and 1 when a disk is present.   (In halos A4 and A8, there are no disky snapshots from $z=2.5-3$.)  All four systems become highly stellar-dominated in the central kpc by $z=1$.  \textbf{Bottom:} Velocity profiles in the simulated galaxies, averaged over the same redshift ranges.  Black lines show the circular velocity $v_c = \sqrt{GM_{\rm enc}/r}$ calculated directly from the density profiles above, while red lines show the actual azimuthal velocity $\bar{v}_\phi$ of the cold gas in the disk.  The velocity dispersion of the gas in the $r$-direction $\sigma_{v_r}$ is shown in blue, with typical values of $\sim$100 km/s.}
  \label{fig:allprofs}
\end{figure*}

We examine the importance of these physical effects on the inference of mass from the rotation curves of gas in disk galaxies using high-resolution cosmological zoom-in simulations run with the Feedback In Realistic Environments (FIRE)\footnote{See the FIRE project web site: http://fire.northwestern.edu.} galaxy formation model \citep{Hopkins2014, Hopkins2018}.  The FIRE simulations capture the important cosmological context, including time-varying inflows and mergers \citep[e.g.,][]{Muratov2015, Angles-Alcazar2017a}, while also achieving high mass resolution on the order of $\sim10^4$ \Msun~per particle. 
This mass resolution allows the simulations to capture all of the main physical processes mentioned above.  Importantly for this study, the simulations resolve the multi-phase ISM and the gas kinematics include the effects of a detailed model for stellar feedback implemented on the scale of individual star-forming regions (see \S \ref{ssec:sims}). 
In the main text, we focus on the case study of massive high-redshift galaxies ($M_* > 10^{11}$ \Msun~at $z>1$).  Our simulations do not include a model for feedback from supermassive black holes (SMBHs), which is likely to modify the dynamics of the galaxy centers and elevate central densities and rotational velocities.  Therefore, we also include in our analysis galaxies down to Milky Way mass at $z=0$, which would be less affected by the absence of SMBH feedback. 
This allows us to verify that our main conclusions apply to galaxies covering a wide range of physical properties. 
In summary, the high resolution and detailed ISM and feedback physics in these simulations enables us to quantify the effects of turbulent pressure support, complex gravitational potentials, and eccentric motion on the rotation curves of the gaseous disks. 

We show that we are generally able to account for most of the difference between the gas rotation velocity and $\sqrt{GM_{\rm enc}/r}$ in disk galaxies as a sum of the physical effects discussed above (including highly turbulent, high-redshift systems).  
These dynamical effects apply to all measurements of gas rotation curves. 
In addition to these physical effects, observational studies are subject to possible biases arising from observational limitations, such as the ability to accurately measure radial profiles of gas surface density and velocity dispersion. 
These observational effects should also be considered, but they are in general functions of the instrument (e.g., spatial resolution) and observational tracers used (e.g., emission lines from HII regions vs. CO emission). 
We do not consider these observational effects in this paper, focusing instead on the underlying physical quantities and forces at play.

The structure of the paper is as follows.  In Section~\ref{sec:methods}, we describe our zoom-in simulations and our methods for calculating the intrinsic kinematic quantities and corrections.  In Section~\ref{sec:results}, we show the physical properties of the simulated high-redshift massive galaxies in several redshift regimes, their rotation curves, and radial profiles of the corrections derived in the previous section.  In Section~\ref{sec:discussion} we discuss the implications of the results on observational studies and the theoretical and observational uncertainties involved, and finally we summarize our findings and conclude in Section~\ref{sec:conclusion}.  We show the results of the same analysis for Milky-Way-like galaxies at low redshift in the Appendix. 

The simulations and analysis herein assume a standard flat $\Lambda$CDM cosmology with $\Omega_{\rm m }=0.32$, $\Omega_{\Lambda}=1-\Omega_{\rm m}$, $\Omega_{\rm b}=0.049$, and $H_{0}=67$ km s$^{-1}$ Mpc$^{-1}$ \citep[e.g.,][]{Planck2018}.
Throughout this paper, distances and velocities are quoted in physical (rather than comoving) units.

\section{Simulations and Methods}
\label{sec:methods}

\subsection{Description of simulations}
\label{ssec:sims}

The FIRE simulations are well-suited to studying the physical effects which modify the rotational velocities in gaseous disks. The zoom-in approach preserves the cosmological context of the halos in which the galaxies reside, while simultaneously providing high enough resolution to capture the  multi-phase ISM, star formation clustering, and time-dependent feedback which are all important to accurately exciting ISM turbulence.
The simulations were evolved using the GIZMO\footnote{http://www.tapir.caltech.edu/~phopkins/Site/GIZMO.html} meshless finite mass (MFM) hydrodynamics solver \citep{Hopkins2015} and the FIRE-2 galaxy formation model \citep{Hopkins2018}. 
The simulations include star formation in dense, self-gravitating gas and stellar feedback driven by radiation pressure, photo-ionization, photo-electric heating, stellar winds (both O-star and AGB), and supernovae (Types I \& II). 
In the main text we focus on the high-redshift massive galaxy regime ($M_* > 10^{11}$ \Msun~at $z > 1$), but in an Appendix we also include examples of an identical analysis performed on simulations down to $z=0$ and Milky Way mass.

The four simulations we examine here are the same discussed by \citet{Angles-Alcazar2017} in the context of supermassive black hole growth and used by \citet{Cochrane2019} to perform a detailed study of their spatially-resolved dust continuum emission. Each simulation was selected from a dark-matter-only volume to have a halo mass of $\sim 10^{12.5}$ \Msun~at $z=2$, and populated with baryons at a mass resolution of $m_{\rm b}=3.3 \times 10^4$ \Msun~(with dark matter particle masses of $m_{\rm DM}=1.7 \times 10^5$ \Msun).  Their initial conditions are the same as halos A1, A2, A4, and A8 in the MassiveFIRE simulation suite \citep{Feldmann2016, Feldmann2017}, but they were evolved with the FIRE-2 model instead of FIRE-1. Supermassive black hole growth was followed passively in these simulations, i.e. feedback from active galactic nuclei (AGN) was not included.
The simulations were evolved down to $z=1$.  The force softening lengths of the particles are 7 pc for stars and black holes and 57 pc for dark matter.  Force softening for the gas is adaptive with the size of the cells, with a minimum of 0.7 pc.  

The stellar mass evolution of these systems from $z=3$ to $z=1$ can be seen in Figure~\ref{fig:massevol}.  Though all four reach stellar masses of $2-5 \times 10^{11}$ \Msun~by $z=1$, they form at different times and in different ways.  In particular, the galaxies in halos A1 and A2 form early, with $\geq 10^{11}$ \Msun~of stars already in place by $z=3$, and experience smooth stellar mass growth thereafter.  Halos A4 and A8, on the other hand, experience more violent assembly histories during the window, gaining much of their stellar mass during mergers.

Images from selected snapshots (marked with black circles in the main panel) are also depicted in Figure~\ref{fig:massevol}.  These show the state of the gas at that time.  In every panel, the brightness of the color scales with the surface density of the gas, while the hue varies according to temperature or line-of-sight velocity.  The galaxies are oriented to be face-on (left) or edge-on (right) according to the angular momentum vector of the cold gas.  In four out of the five snapshots shown, a clear galactic disk is present, marked by a thin shape and dipole in line-of-sight velocity when viewed edge-on.  The sizes and thicknesses of the disks vary with time and between simulations, and each simulation experiences a series of episodes of stable disks, unstable periods, and disk reformation.   The images on the lower left, for example, depict a snapshot where there is not an organized disk present; rather there is a cavity around the galaxy following an outburst from stellar feedback.  These images are an indication of the variety of disks (and non-disks) that occur throughout the course of the simulations.

\begin{figure*}
  \centering
  \includegraphics[width=2.1\columnwidth]{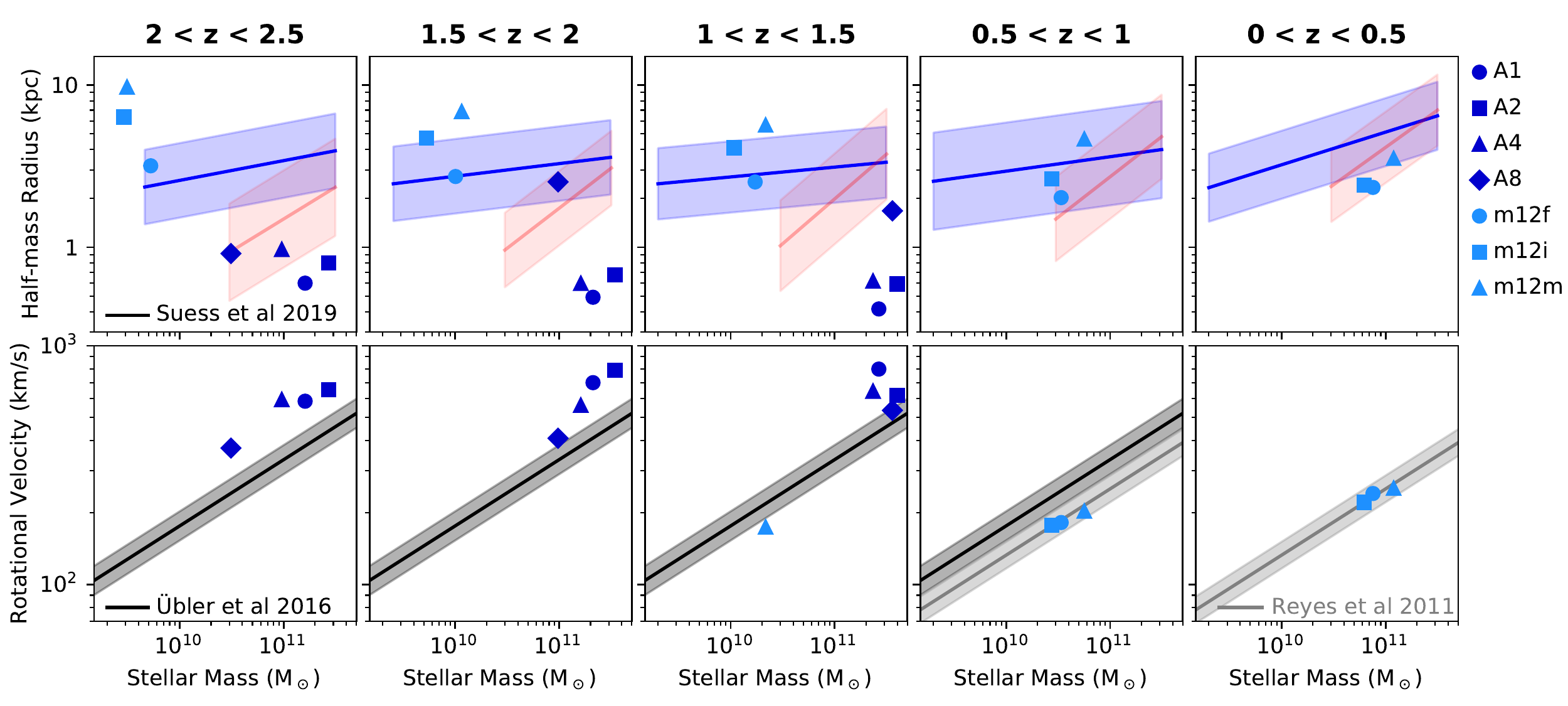}
  \caption{Size-mass (top) and Tully-Fisher (bottom) relations of our simulated galaxies (points) averaged over five redshift intervals, with the observed relations (lines) for context.  In the top row, size $R_{1/2}$ is measured from the simulations as the projected stellar half-mass radius with the galaxy face-on. (For a more thorough analysis of the sizes that would be observationally estimated for the simulated massive galaxies, see Parsotan et al, in prep.)  The simulated massive galaxies are significantly more compact than expected for star-forming galaxies given the observed relation \citep{Suess2019b} and scatter \citep{Mowla2019}, shown in blue.  (They are closer to, but still more compact than, the relation for quiescent systems shown in red.)  In the bottom row, rotational velocity is measured for disky snapshots as the intrinsic $v_\phi$ of the cool gas at a radius $r=2R_{1/2}$.  The over-compactness of the stellar distribution in the massive systems produces higher rotational velocities than observationally expected, such that the simulated massive galaxies lie well above the observed high-$z$ Tully-Fisher relation \citep{Ubler2017}.  The Milky-Way-mass simulated galaxies (m12X, discussed in the Appendix) have more extended sizes consistent with the observed sizes and the local Tully-Fisher relation \citep{Reyes2011}.}
  \label{fig:sizemass}
\end{figure*}

Mass density profiles of the stars, gas, and dark matter in these four galaxies during disky phases are shown in Figure~\ref{fig:allprofs}.  Each panel contains 3-4 lines for each mass component representing averages over the redshift ranges $z$ = 2.5 - 3 (if a disk was already present), 2 - 2.5, 1.5 - 2, and 1 - 1.5.  All four simulations are highly stellar-dominated in the central kpc by $z=1$.  
The two systems which are already in place at $z=3$ (halos A1 and A2) experience very little change in their dark matter halo profiles during this time, while the dynamic growth of halos A4 and A8 appears in all three mass components.  All four simulations retain a significant amount of star-forming gas throughout most epochs of their formation, though they often experience episodes of outburst from stellar feedback that can deplete their nuclear gas reservoirs \citep{Angles-Alcazar2017}.

The strong baryon domination in the central region is qualitatively consistent with observed galaxies of similar masses and redshift \citep[see][and references therein]{Genzel2017}.  However, the simulated massive galaxies appear even more compact than most observed galaxies in this regime. 
In Figure~\ref{fig:sizemass}, we show the average projected stellar half-mass radii ($R_{1/2}$) of these galaxies as dark blue points over several redshift ranges, measured by rotating the galaxy face-on, calculating the projected stellar mass profile out to 30 kpc, and finding the radius which enclosed half of the total mass.\footnote{In addition to this direct measurement, we also performed single- and double-S\'ersic fits to the stellar surface density profiles.  We found that these typically recovered reasonable matches to the true projected half-mass radii and that for most snapshots, the double-S\'ersic form provided a good fit to the surface density profile.  The best-fit profiles typically had an inner ``bulge" with S\'ersic index $n_{\rm bulge} \sim 2-4$ and effective radii $R_e \sim 0.5-2$ kpc, plus an outer ``disk" ($n=1$) with scale radius $R_d \sim 5-10$ kpc.  Requiring a single component yielded less-good but still-reasonable fits with $R_e \sim 0.4-2$ kpc and high $n \sim 4-8$.}  This measurement of intrinsic size can be reasonably compared to the observational data described below, but does not involve forward modeling and so neglects some of the complications of deriving sizes from observational data.  For a more thorough analysis of the size-mass relation of these simulated massive galaxies which does include forward modeling, see Parsotan et al, in prep.

For comparison, the mean relations between stellar mass and half-mass radius estimated from observational data by \citet{Suess2019a,Suess2019b} are shown as blue and red lines for star-forming and quiescent galaxies respectively.  \citet{Suess2019a} obtain these sizes by adjusting the observed light profiles measured in elliptical annuli with the radius-dependent mass-to-light ratio implied by observed radial color gradients, producing estimated mass surface density profiles whose half-mass radii can be compared to simulation data without requiring forward modeling.  The scatter in the observed relations from \citet{Mowla2019} is shown in shaded regions around the mean lines.

Our simulated massive galaxies typically have sub-kpc half-mass radii, which is significantly more compact than expected in comparison to the observed size-mass relation for star-forming galaxies.  Although the galaxies are star-forming throughout the duration of the simulations, their sizes even lie below the relation for quiescent galaxies.  This is most likely due to the absence of AGN feedback in the simulations, which allows too many baryons to condense and form stars near the center and does not lead to galaxy quenching. 
We note that similarly compact stellar radii are found in other cosmological zoom-in simulations \citep[][]{Choi2018}. 
Although \cite{Choi2018} show that including AGN feedback increases stellar sizes in massive galaxies, their particular AGN feedback model appears to still under-predict observed stellar effective radii by a factor $\sim2$ on average.

The over-compactness of the stellar distribution in the simulated galaxies has a direct effect on the magnitude of the rotational velocities, shown in the Tully-Fisher relation in the bottom panels of Figure~\ref{fig:sizemass}.  Here, rotational velocity is measured from disky snapshots as the intrinsic $v_\phi$ of the gas at a radius $r=2R_{1/2}$.  For comparison, the black line shows the stellar Tully-Fisher relation from \citet{Ubler2017} measured from galaxies at $0.6 \leq z \leq 2.9$.   The relation measured by \citet{Ubler2017} is offset from (higher than) the local relation, so in the two lowest-redshift panels we also show the local relation as measured by \citet{Reyes2011} from galaxies at $z < 0.1$.  Because the simulated massive galaxies are so centrally concentrated, their rotational velocities at $2R_{1/2}$ are much higher than expected, which places them well above the observed Tully-Fisher relation.

A sample of lower-mass simulated galaxies (m12X) which reach Milky Way mass by $z=0$ are also shown in light blue points in Figure~\ref{fig:sizemass}.  These systems have more extended stellar sizes which are more consistent with the observations and result in rotational velocities which place them on the local Tully-Fisher relation.  Fortunately, our analysis of these lower-mass simulations \citep[which have been shown to be in good agreement with observed $\sim L^*$ galaxies, e.g.][]{Wetzel2016, El-Badry2018b} in the Appendix indicates that our main results apply over a wide range of galaxy properties and are not specific to our more massive galaxy simulations.  We therefore defer the question of how to best model AGN feedback and its effects on galaxy properties to future work.  In this paper we focus instead on an analysis of the different effects on gas rotation curves which can be identified in the self-consistent dynamical models provided by the present simulations.

\subsection{Kinematic measurements and corrections}
\label{ssec:calcs}

In the kinematic analysis that follows, we are careful to select only those snapshots where a well-organized gaseous disk is present, much as an observational study would select only systems with clean and smooth gas velocity profiles (see e.g. \citet{Wisnioski2015}).  Specifically, we only use snapshots where the disk is (i) rotation-dominated, i.e. $\bar{v}_\phi/\sigma > 4$ for the gas at twice the stellar half-mass radius, and (ii) smoothly distributed.  To enforce the latter, at each snapshot we fit the azimuthally-averaged surface density profile to a double power law, which is a flexible enough functional form to find a good fit whenever a smooth disk is present.  The typical slopes are between 0 and -1 in the inner part of the disk and between -2.5 and -4.5 in the outer part of the disk.  We discard snapshots with positive inner slopes or with very large $\chi^2$ fit values indicating gaps, jumps, or other non-equilibrium situations.  (We also fit exponential profiles, but it is a less flexible functional form which does not provide as good a fit - see e.g. Figure~\ref{fig:posterchild}.)  Snapshots which meet these two criteria (approximately 40\% of all snapshots from $z=1-3$) are marked with stars in Figure~\ref{fig:massevol}.

For all calculations, we first define the galaxy's rotation axis to align with the angular momentum vector of the cool ($10^{3.5}$ K $< T < 10^{4.5}$ K) gas within 10 kpc of the galaxy center.  Then, in annuli equally spaced in $\log r$, we calculate the scale height $z_h(r)$ of the cool gas where $\sum_{|z_i| < z_h} m_i / \sum_i m_i = 1 - 1/e$ assuming an exponential vertical profile, and select the cool gas particles which lie between $\pm z_h$.  The temperature cut was chosen for the purpose of selecting the gas which fills the disk and whose turbulence is responsible for supporting it (rather than selecting the cold clouds embedded within or the hot outflowing gas), but in practice we find that the results are not very sensitive to this choice -- a selection of all gas below $10^{4.1}$ K, for example, yields a similar outcome.  In the following analysis, we compare the rotational velocities of the gas thus selected with the spherically-symmetric, zero-pressure assumption that $\bar{v}_\phi = \sqrt{GM_{\rm enc}/r}$ as calculated from the density profiles shown in Figure~\ref{fig:allprofs}.  We then discuss the physical effects that lead to deviations from this assumption.  

\subsubsection{Intrinsic velocities}
\label{sssec:intrinsicv}

To calculate the true rotational velocity profile in the disk, we select cool gas in logarithmically-spaced radial bins as described above, and find the average velocity in the $\phi$-direction \vphi over the gas particles in each annulus.  We also calculate the radial velocity dispersion profile $\sigma_{v_r}(r)$ by taking the standard deviation of the distribution of the radial velocity $v_r$ over the same selection of gas resolution elements.  (In general we find that $\sigma_{v_r}$, $\sigma_{v_z}$, and $\sigma_{v_\phi}$ typically have similar values, but $\sigma_{v_r}$ is directly related to the radial turbulent pressure gradient we describe below.)

The velocity profiles thus compiled are shown in the bottom row of Figure~\ref{fig:allprofs}, with $\sqrt{GM_{\rm enc}/r}$ in black, \vphi in red, and $\sigma_{v_r}$ in blue.  Both the ``expected" and measured rotational velocities can be quite high -- often exceeding 500 km/s -- due to the extremely compact stellar mass distributions.  The velocities are most extreme in the galaxies which formed earliest.  We find one-dimensional radial velocity dispersions of about 100 km/s which are nearly constant with radius except for a peak in the very center.  As discussed in the previous section, the central densities and rotational velocities are higher in the simulations than observed in real galaxies, likely due to the absence of AGN feedback.  As such, we caution the reader that the exact values of, e.g., the central rotational velocities and stellar densities should not be considered quantitative predictions for the properties of massive high-redshift galaxies.  A common pattern that emerges in these profiles is that there is a region, typically from 1-3 kpc, where \vphi is in agreement with $\sqrt{GM_{\rm enc}/r}$, but that \vphi falls away at both large and small radii.  

\subsubsection{Turbulent pressure correction}
\label{sssec:presscalc}

One physical effect which can reduce rotation is support from radial gradients in the turbulent pressure of the disk (see \citet{Burkert2010} for a discussion).  This appears as an additional term in the force balance
\begin{equation}
\frac{\bar{v}_\phi^2}{r} = f_g(r) + \frac{1}{\rho}\frac{dP}{dr}
\end{equation}
where $f_g(r)$ is the radial component of the specific force of gravity at $r$, $\rho$ and $P$ are the density and pressure of the gas, and the left-hand side of the equation assumes circular orbits.

If turbulence is the dominant source of pressure (i.e. thermal pressure is negligible, which is the case for the cool gas in massive high-z systems), $P = \rho \sigma^2$ (where $\sigma$ is the one-dimensional velocity dispersion) and 
\begin{equation}
\frac{\bar{v}_\phi^2}{r} = f_g(r) + \frac{\sigma^2}{r} \frac{d \log (\rho \sigma^2)}{d\log r}. 
\label{eqn:press}
\end{equation}
To measure this force on the gas from the radial turbulent pressure gradient, we first calculate the pressure profile $P(r) = \sum_i m_i (v_{r,i} - \bar{v}_r(r))^2 / (2 \pi r ~ \Delta r ~ 2z_h(r))$ for the cool gas particles in each annulus, where $m_i$ and $v_{r,i}$ are the mass and radial velocity of individual particles and $\bar{v}_r(r)$ is the average radial velocity within the annulus (which is close to zero except at very large radii).  We then measure the slope $d \log P / d \log r$ using a first-order finite difference approximation.  

Here, we are able to measure this support from turbulent pressure gradients (also known as ``gas asymmetric drift") directly using the full simulation data, but observational studies must attempt to estimate it using various observable proxies - see Section~\ref{ssec:pressurediscussion} for a discussion of how these estimates differ.

\subsubsection{Gravitational potential correction}
\label{sssec:gravcalc}

An additional correction to the velocity arises from the fact that these systems are not spherically symmetric, as the equation $\bar{v}_\phi = \sqrt{GM_{\rm enc}/r}$ assumes.  Rather, they are composed principally of an ellipsoidal stellar bulge and dark matter halo and a flat gaseous and stellar disk, in addition to substructure such as satellite subhalos.  Observationally, this is typically accounted for by fitting a bulge-disk decomposition to find the relative mass of each component and combining the potentials produced by spherical and disky distributions of mass. 

With simulation data, however, we can directly calculate the specific gravitational force (i.e., the acceleration) felt by a test particle lying on the disk midplane.  We perform such a calculation by brute force, summing the gravitational forces exerted by the individual mass resolution elements on test particles at four locations within each radial annulus.  All mass within 100 kpc of the test particle is included in this calculation.  For efficiency, we downsampled the resolution elements by a factor of 10, which is sufficient for the calculation to converge.  The numerical effect of gravitational softening, which occurs at typical scales of 1, 7, and 57 pc for gas, stars, and dark matter respectively, was included in the force calculation.  We then average the $r$-component of the specific force at each of the four locations to find $f_g(r)$.  This specific force will differ from $GM_{\rm enc}/r^2$, i.e.
\begin{equation}
f_g(r) = \frac{GM_{\rm enc}}{r^2} + \delta f_g(r)
\end{equation}
where $\delta f_g(r)$ represents the deviation from spherical symmetry.  Incorporating this definition into the force balance,
\begin{equation}
\frac{\bar{v}_\phi^2}{r} = \frac{GM_{\rm enc}}{r^2} + \delta f_g(r) + \frac{\sigma^2}{r} \frac{d \log (\rho \sigma^2)}{d\log r},
\end{equation}
or, in terms of mass,
\begin{equation}
\frac{\bar{v}_\phi^2 r}{G} = M_{\rm enc} + \frac{r^2~\delta f_g(r)}{G} + \frac{\sigma^2r}{G} \frac{d \log (\rho \sigma^2)}{d\log r}.
\end{equation}
We define each of these terms as
\begin{align}
M_{\bar{v}_\phi} & = \frac{\bar{v}_\phi^2 r}{G} \label{eqn:Mvphi} \\
\Delta M_{\rm grav} &= -\frac{r^2~\delta f_g(r)}{G} \label{eqn:dMgrav} \\
\Delta M_{\rm press} &= -\frac{\sigma^2r}{G} \frac{d \log (\rho \sigma^2)}{d\log r}. \label{eqn:dMpress}
\end{align}
such that
\begin{equation}
M_{\rm enc} = M_{\bar{v}_\phi} + \Delta M_{\rm grav} + \Delta M_{\rm press}. \label{eqn:massbalance}
\end{equation}
Throughout this paper, we measure the terms on the right-hand side of this equation and compare them to the true enclosed mass $M_{\rm enc}$ to examine the importance of various physical effects on the estimation of dynamical mass from observed rotation.

\subsubsection{Measuring non-circularity}
\label{sssec:noncirc}

\begin{figure}
  \centering
  \includegraphics[width=\columnwidth]{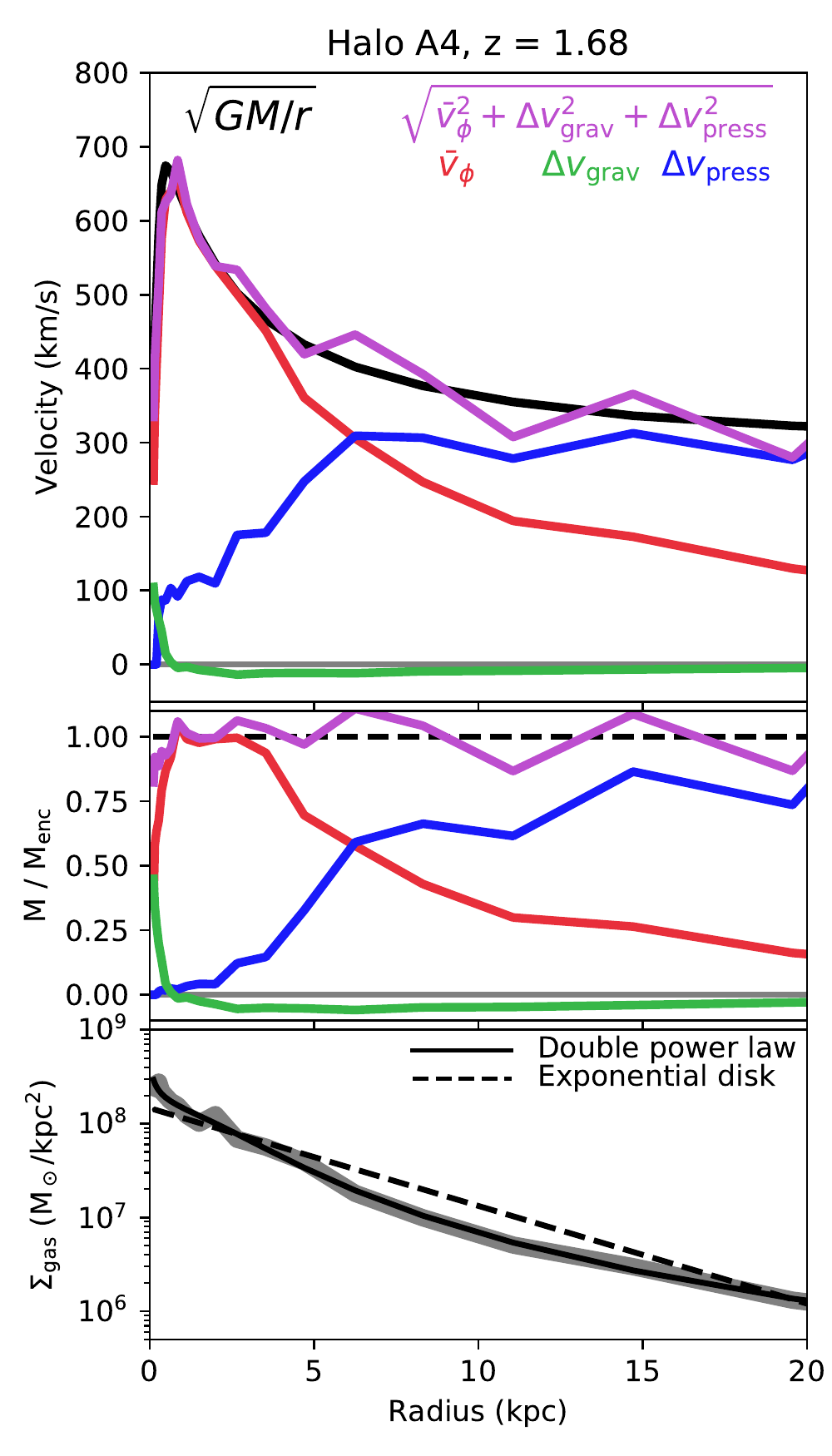}
  \caption{Gas rotation curve and corrections for a single snapshot of halo A4 at $z=1.68$.  The intrinsic rotation $\bar{v}_\phi$ (red) of the gas as well as the circular velocity $\sqrt{GM_{\rm enc}/r}$ (black) are the same as shown in Figure~\ref{fig:allprofs}.  The corrections due to support from turbulent pressure gradients in the disk ($\Delta v_{\rm press}$, blue) and the non-sphericity of the gravitational potential ($\Delta v_{\rm grav}$, green) add in quadrature with $\bar{v}_\phi$ to produce a ``corrected velocity" (purple) which approximately recovers $\sqrt{GM_{\rm enc}/r}$.  The middle panel shows the same four lines in terms of $\Delta M$ relative to the true enclosed mass, as defined in Equations~\ref{eqn:Mvphi}-\ref{eqn:massbalance}. The bottom panel shows the surface density profile of the cool gas, where the thick grey line is the data from the simulation and the black lines are the best fits to a double power law (solid) or exponential disk (dashed) form.}
  \label{fig:posterchild}
\end{figure}

Finally, there are numerous physical circumstances that produce non-circular orbits in the gaseous disk.  Our simulated disks often produce non-axisymmetric structures, such as bars or other highly elliptical orbits, in the central regions.  Such structures can be identified with careful analysis of kinematic maps (see e.g. \citet{Oman2017}), but can be difficult to spot if unresolved, as may be the case at high redshift.  In other places in the disk, the gas may not be in equilibrium - e.g. if there are radial inflows or outflows due to feedback.  We find that the interior region of the disk is likely to be susceptible to these effects, as it experiences volatility from feedback.  The exterior regions of the disk are also vulnerable to perturbations from inflows. 

\begin{figure*}
  \centering
  \includegraphics[width=1.6\columnwidth]{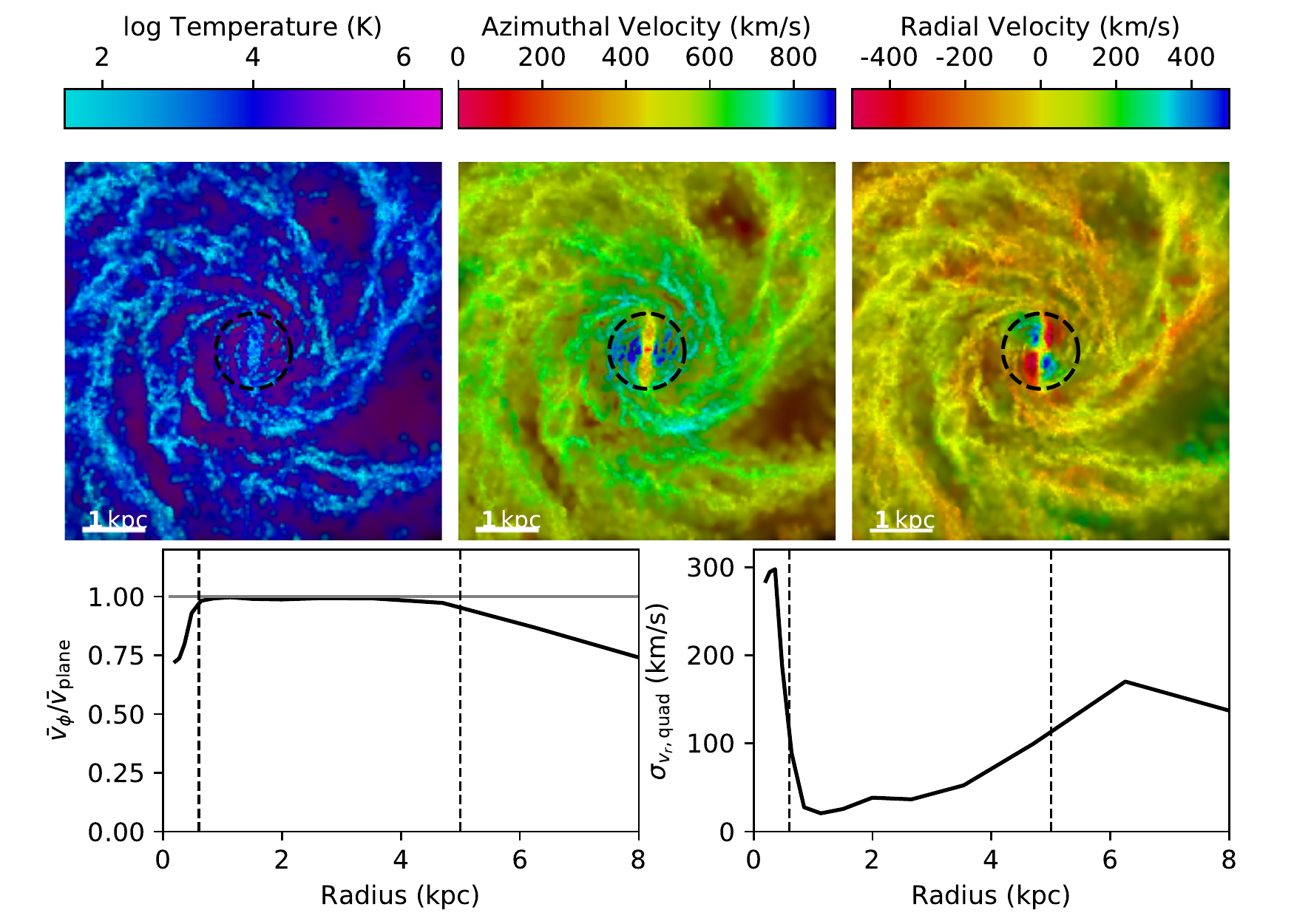}
  \caption{Maps of the gaseous disk during a snapshot when the inner kpc is dominated by a bar.  In each map, the brightness indicates the gas surface density and the hue indicates the temperature (left), azimuthal velocity $v_\phi$ (center), or in-plane radial velocity $v_r$ (right).  In the velocity maps, a clear non-axisymmetry is present which demonstrates that the gas is on highly elliptical orbits.  Black dashed lines indicate radii of 0.6 kpc, within which the bar begins to dominate the rotational motion, and 5 kpc, beyond which the inflowing gas has not yet circularized.  The two panels below quantify orbital axisymmetry. The left panel shows a radial profile of the fraction of in-plane velocity $\left<\sqrt{v_\phi^2 + v_r^2}\right>$ contributed by $\bar{v}_\phi$.  The right panel shows the dispersion between the $\bar{v}_r$ values measured in four quadrants: $\phi = 0 - \pi/2, \pi/2 - \pi, \pi - 3\pi/2$, and $3\pi/2 - 2\pi$.  These quantities clearly deviate inside the dashed circle, and are used in Section~\ref{sssec:noncirc} to define regions in the disk where the orbits are approximately circular.}
  \label{fig:acirc}
\end{figure*}

To isolate the region of the disk where the gas is orbiting in a circular fashion (i.e. where $\bar{v}_\phi$ can be meaningfully compared to $\sqrt{GM_{\rm enc}/r}$), we apply the following criteria: (a) that the in-plane motion is dominated by rotation, $\bar{v}_\phi/\left<\sqrt{v_\phi^2 + v_r^2}\right> >$ 0.9, (b) that there are no significant radial inflows or outflows, $|\bar{v}_r| <$ 85 km/s, and (c) that the motion is uniform within the annulus.  To check the latter, we measure $\bar{v}_r$ in quadrants of $\pi/2$ and calculate the dispersion among the quadrants, requiring $\sigma_{v_r, \rm quad} <$ 100 km/s.  In Figure~\ref{fig:acirc}, we show one example of the complexity of the velocity fields in the interior of the disk and the corresponding radial profiles of these quantities (see Section~\ref{ssec:noncircresults}).

\section{Results}
\label{sec:results}

\subsection{Velocity structure in high-z massive disks}

When the massive galaxies form stable disks at high redshift, they can be highly rotation-dominated despite their large turbulent velocity dispersions. 
The rotation curves of our simulated massive galaxies (as measured by intrinsic $\bar{v}_\phi$) peak at velocities from 300-800 km/s, extremely fast rotation which is required by the very dense stellar distribution at the galaxy centers.  We find that it is common for the galaxies to have a structure with significant velocities in $v_r$ or $v_z$ in the central kpc (see Figure~\ref{fig:acirc}). In particular, halo A1 possesses a misaligned inner disk which persists for $\sim$1.4 Gyr (visible in the rightmost images in Figure~\ref{fig:massevol}), and all four galaxies possess a bar for a significant fraction of the simulation duration.  Other than A1's inner misalignment, the disks are generally planar without long-standing warps - all the rotational velocities we show in this paper assume that the global galactic angular momentum vector applies to all radii, but we find that defining it separately for each annulus yields essentially the same results.  The gaseous disks can be relatively compact (though rarely as compact as the stars) with some simulations displaying dynamic fluctuations in disk size from $\sim$kpc to $\sim$10 kpc scales (or the reverse).  

We show one example of the intrinsic velocity profile during a single snapshot of halo A4 at $z=1.68$ in Figure~\ref{fig:posterchild}.  In this particular snapshot, the rotational velocity peaks at 650 km/s, and matches $\sqrt{GM_{\rm enc}/r}$ from $r$ = 0.8 - 3 kpc.  The one-dimensional radial velocity dispersion is about 100 km/s at all radii (see Figure~\ref{fig:allprofs}). 

In the bottom panel, we also show the surface density profile of the cool gas during this snapshot in grey.  Overlaid are two fits to this profile: an exponential disk $\Sigma \propto e^{r/R_d}$ (dashed line), and a double power law form (solid line).  We find that in most cases, as in this particular example, the double power law provides a better fit to the data.  We find that exponential disk fits give scale radii $R_d \sim$ 1.5 - 5 kpc and double power law fits give similar break radii, $R_b \sim$ 2 - 6 kpc.  The break or scale radii signify the location in the disk where the slope of the surface density profile steepens, and hence where the pressure term becomes larger.

We have chosen to show this snapshot because it clearly exemplifies the physical trends that govern our simulated disks on average, but we note that individual snapshots are typically noisier than this one.  In the subsequent subsections, we show how the physical effects we measured have modified the rotational velocity for this snapshot, then show general results for all four simulations in four redshift regimes (Figure~\ref{fig:tileresults}).

\subsection{Effect of pressure gradient}

The simulated massive galaxies have high one-dimensional gas velocity dispersions of 100-150 km/s which are approximately constant throughout the disk.  These high velocity dispersions mean that there is the potential for strong radial support by pressure gradients.  Two physical quantities affect the magnitude of this correction: the velocity dispersion, and the density gradient (see Equation~\ref{eqn:press}).  We find that surface density is well-represented by a double power law, with a shallow slope in the main part of the disk (typically between 0 and -1) and a steepening in the outer disk (typically between -2.5 and -4.5).  In the main part of the disk where the gas surface density profile is shallow, the pressure correction is small compared to $\bar{v}_\phi$.  As the profile steepens in the outer disk, however, the pressure gradient becomes an increasingly important source of support and can suppress the rotational velocity by 10-40\%, leading to an underestimate of the mass if not accounted for.

\begin{figure*}
  \centering
  \includegraphics[width=1.8\columnwidth]{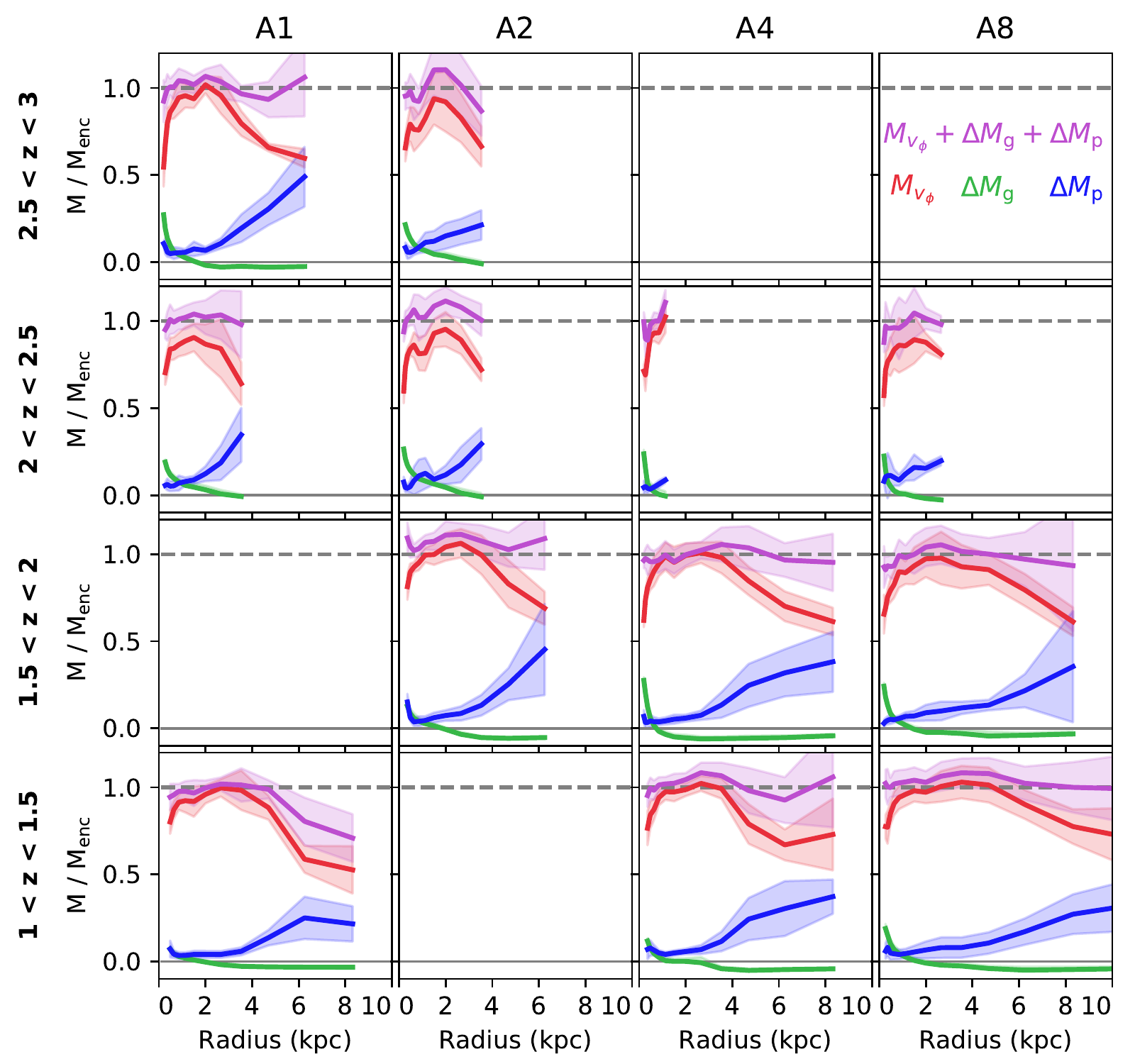}
  \caption{Mass estimation relative to true enclosed mass $M_{\rm enc}$, where each term is defined as in Equations~\ref{eqn:Mvphi}-\ref{eqn:dMpress}.  In each panel, red lines show the mass inferred directly from rotation $M_{v_\phi}$, blue lines show the correction accounting for radial gradients in turbulent pressure $\Delta M_{\rm press}$ (shown as $\Delta M_{\rm p}$), green lines show the correction accounting for the non-spherical gravitational potential $\Delta M_{\rm grav}$ (shown as $\Delta M_{\rm g}$), and purple lines show the combination of the three $M_{v_\phi} + \Delta M_{\rm grav} + \Delta M_{\rm press}$.  Solid lines represent the median values for a single simulation (varying by column) over a redshift range $\Delta z = 0.5$ (varying by row).  Only snapshots which meet the disk criteria described in Section~\ref{ssec:calcs} and regions of the disk which meet the circularity criteria described in Section~\ref{sssec:noncirc} are included.  Shaded regions indicate the 25th-75th percentile of the variation among snapshots.  Including these cuts and corrections allows the recovery of $M_{\rm enc}$ on average, demonstrating that these are the dominant physical effects which cause $M_{v_\phi}$/$M_{\rm enc}$ to differ from unity.}
\label{fig:tileresults}
\end{figure*}

The shape and magnitude of the pressure term $\Delta M_{\rm press}$ at $z=1.68$ in halo A4 is shown in blue in Figure~\ref{fig:posterchild}.  At this moment, the pressure correction is negligible at small radii but is increasingly important outside of 3 kpc and dominates outside of 6 kpc.  Radial profiles of the magnitude of the pressure term relative to $M_{\rm enc}$ are shown in blue in Figure~\ref{fig:tileresults} for all simulations.  The pattern of the increasing importance of turbulent pressure gradients in the outer disk appears generically in all four simulations at all redshifts where an organized disk is present, though the radius at which the pressure term becomes significant differs.  The importance of this effect in the outer disk seems to be independent of redshift, reducing the estimated mass by 10-40\% up to $z\sim 1$ (when the simulations were halted).  At high redshift ($z>2$) the turbulent pressure gradient can also be significant within the main body of the disk, with $\Delta M_{\rm press}/M_{\rm enc} \sim 10\%$ even at small radii in several simulations.  In these cases, the mass inferred directly from rotation $M_{v_\phi}$ underestimates the true mass throughout the entire disk.

\subsection{Effect of non-spherical potential}

In our simulated galaxies, the dominant mass component within the galaxy itself is the stellar distribution, which comprises about 75\% of the mass within 2 kpc on average.  The stellar bulge and disk are therefore primarily responsible for determining the shape of the gravitational potential.  The bulge fraction of the stellar component in these simulated galaxies (measured kinematically as $2 \times$ the fraction of stars which are counter-rotating with respect to the gaseous disk) is around 40\%, so about 60\% of the stars can be expected to reside in a rotating, non-spherical ``disk."  This oblate stellar disk will alter the potential, which we  measure by brute force as described in Section~\ref{sssec:gravcalc}.

The green line in Figure~\ref{fig:posterchild} shows the correction to the mass estimate accounting for a non-spherical gravitational potential $\Delta M_{\rm grav}$ for halo A4 at $z = 1.68$.  Outside of 1 kpc, this correction is negligible (and in fact slightly negative).  The innermost region can be affected by the asphericity of the matter distribution, however the true magnitude of this effect likely depends on how AGN feedback affects central densities.  As in the previous section, this pattern holds for all four simulations in all redshift bins (see Figure~\ref{fig:tileresults}).

\subsection{Effect of non-circular orbits}
\label{ssec:noncircresults}

Additional discrepancies between corrected velocity and $\sqrt{GM_{\rm enc}/r}$ can largely be attributed to non-circular motions caused either by non-axisymmetric structures or non-equilibrium situations (i.e. inflows or outflows).  Figure~\ref{fig:acirc} shows one such example of a disk with a strong central bar.  A face-on map of gas density and temperature is shown in the upper left panel, with brighter colors indicating denser regions and bluer colors indicating colder regions.  The two panels to the right show the velocity structure in the disk, with $v_\phi$ mapped in the center and $v_r$ mapped on the right.  An overall radial gradient in $v_\phi$ is visible, but a bar disturbs the gradient in the central region.  This bar appears as a quadrupole in $v_\phi$, as the orbits are strongly elliptical.  Here, $v_\phi$ alone becomes a poor indicator of total rotational support.  Such regions with significantly non-circular orbits should in general be excluded from dynamical mass analyses.

The elliptical motion is quantified in the bottom two panels of Figure~\ref{fig:acirc}. The bottom left panel shows the profile of the average azimuthal velocity relative to the average in-plane velocity, $\left<\sqrt{v_\phi^2 + v_r^2}\right>$.  While the in-plane velocity is wholly dominated by $v_\phi$ outside the barred region, within it $v_r$ becomes a significant fraction of the velocity.  We also examine the uniformity of radial motion throughout the annulus by measuring $\bar{v}_r$ in four quadrants of $\Delta \phi = \pi/2$ and comparing their values.  In regions with strongly elliptical orbits, the standard deviation of these four values $\sigma_{v_r, \rm quad}$ will be high.  In this particular example, $\sigma_{v_r, \rm quad}$ exceeds 100 km/s in that central region.  Similarly, the outer edge of the disk (beyond which inflows and asymmetries begin to dominate) can be identified using the same quantities.  In this example, the outer edge of the disk is around 5 kpc, marked with a dashed vertical line in the lower panels.  When we calculate modifications to rotational velocity, we first select circularized regions of the disk using these quantities, as described in Section~\ref{sssec:noncirc}.

We find such non-axisymmetric structures commonly in the central regions of the simulated galaxies.  The centers of these disks tend to be volatile, with episodic central outbursts followed by inflows.  The outer regions of the disks are also more likely to be eccentric, as they accrete cool gas on elliptical orbits which have not yet circularized.  In Figure~\ref{fig:tileresults} when we show average results within a given redshift range, we include only the largest contiguous region of the disk which satisfies the circularity criteria.  

On average, when (i) a well-organized disk is present, (ii) the circularized region of the disk is selected, and the effects of (iii) radial gradients in turbulent pressure and (iv) non-spherical gravitational potentials are accounted for, we find that a reasonable estimate of the enclosed mass profile may be recovered from the observed rotation profile.

\section{Discussion}
\label{sec:discussion}

\subsection{High-redshift disk properties}
\label{ssec:diskprops}

It is clear that our simulated massive high-z disks are very different in their physical properties from disks in the local Universe.  
They are more extreme overall, with small sizes and high velocities, and are highly stellar-dominated with dark matter, gas, and stars comprising 22\%, 3\%, and 75\% of the total mass within 2 kpc on average respectively.  The DM fraction within 2 kpc is roughly consistent with the $\sim$20-25\% figures reported by \citet{Lovell2018} and \citet{Teklu2017} for galaxies at analogous stellar mass in the IllustrisTNG and Magneticum Pathfinder large-volume simulations respectively.  Both the stellar and gas distributions are compact ($R_{1/2} \approx$ 1 kpc for stars and 3-6 kpc for cold gas), with a dark matter fraction of only $\sim$8\% within the stellar half-mass radius. 
This small dark matter mass fraction is consistent with observational estimates from e.g. \citet{Genzel2017}.  The compact stellar mass distributions lead to high velocities in both rotation ($\sim$500 km/s) and dispersion ($\sim$100 km/s) in the simulations.

Although observations show that $z \sim 1-2$ massive galaxies are very compact relative to the low-redshift Universe \citep[e.g.,][]{Trujillo2007, VanDokkum2008, Szomoru2010}, in detail our simulated massive galaxies appear somewhat too compact relative to observed galaxies in the same mass range at high redshift, likely due to the lack of AGN feedback.  
If the centers of the galaxies were less dense, we would expect certain other quantitative properties of the system to change as well.  A less-compact bulge would change the gravitational potential and lead to lower peak rotational velocities. 
Additionally, this could imply that the effect of turbulent pressure support on rotational velocities in high-z massive galaxies would actually be \textit{stronger} than we show here \citep[although in some disk models with Toomre parameter $Q\sim1$ the velocity dispersion scales with $v_{\rm c}$, e.g.][]{Thompson2005, FG13}. 

Because some galactic properties may change with the inclusion of AGN feedback, we caution that certain results presented herein (e.g., the magnitude of the peak rotational velocity, the location of the region where $v_\phi \approx v_{\rm circ}$) should not be considered robust quantitative predictions of the properties of high-redshift galaxies.  However, we expect that the general trends presented here will hold when the effects of a reasonable AGN feedback model on galactic structure is included (although not necessarily in periods when AGN feedback is ``on'' and potentially strongly disturbing the ISM).  As a proof of concept, we show in the Appendix that we are able to account for the different effects determining gas rotation velocities not only in extremely compact, massive galaxies at high redshift but also in much more extended low-redshift disks, similar to the Milky Way.  Despite their very different quantitative properties, we find that the same general trends hold for these systems.

\begin{figure*}
  \centering
  \includegraphics[width=1.8\columnwidth]{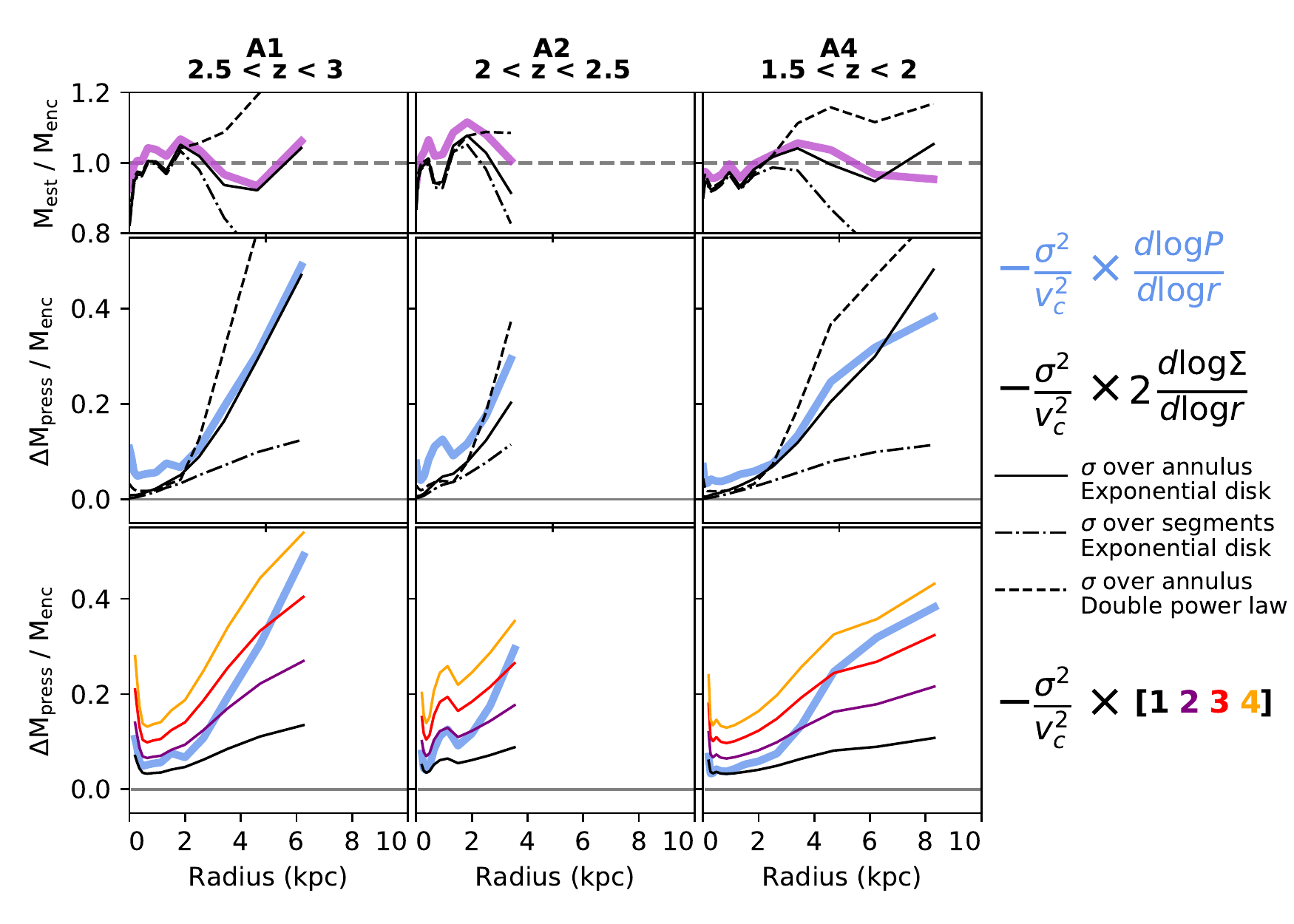}
  \caption{Comparison of the true effect of the pressure gradients on dynamical mass estimates with various quasi-observational means of estimating it.  Each row shows a different simulation and redshift range.  Thick blue and purple lines are a subset of those shown in Figure \ref{fig:tileresults}, and show the true modification of the mass estimate $\Delta M_{\rm press}/M_{\rm enc} = \sigma^2/v_c^2~d\log P/ d\log r$.  Each line is a median over a redshift range $\Delta z = 0.5$.   \textbf{Top and middle:}  Black lines show the observational estimate $\Delta v^2 = -2 \sigma^2~d\log \Sigma/ d \log r$, where $\Sigma$ is gas surface density, with several variants of how $\sigma$ and $\Sigma$ are measured.  Solid lines utilize an exponential disk fit to the surface density and a value of $\sigma$ calculated over an entire annulus.  Dashed lines use the same exponential disk fit, but instead split the annulus into 16 $\phi$-bins before measuring $\sigma$.  Dot-dashed lines replace the exponential profile with a double power law fit to the surface density.
  The middle row shows the estimated correction $\Delta M_{\rm press}$, while the top row shows the total estimated mass relative to the true mass. \textbf{Bottom:} Colored lines show the estimate if the pressure profile has a slope of -1 (black), -2 (purple), -3 (red), or -4 (orange).  The pressure profiles typically have slopes between -1 and -2 in the main body of the disks and steepen to -3 or -4 in the outer regions.  All three quasi-observational measurements underestimate the effect of the pressure gradient within the disk itself, because the surface density profile is typically shallow and its slope is a poor proxy for the slope of the pressure profile.  In the outer disk, the surface density profile steepens and an exponential disk fit can give a good approximation to the pressure component, but estimates which use $\sigma$ measured in small  $\phi$-segments still under-predict the effect.}
  \label{fig:pressureestimators}
\end{figure*}

When we measure the azimuthal velocity profiles $\bar{v}_\phi(r)$ of the gaseous disks for the massive high-redshift galaxies, we find that there is a region from $\sim$1-3 kpc from the galaxy center where $\bar{v}_\phi$ is well-matched to $\sqrt{GM_{\rm enc}/r}$, meaning that the gas rotation curve could be straightforwardly used to accurately recover the enclosed mass.  Outside of this region, however, it falls faster than $\sqrt{GM_{\rm enc}/r}$ would predict.  There, we find that other physical effects become important which can suppress $\bar{v}_\phi$.

\subsection{The importance of turbulent pressure gradients and their observational estimates}
\label{ssec:pressurediscussion}

The presence of a radial gradient in turbulent pressure, indicated with a blue line in Figures \ref{fig:posterchild} and \ref{fig:tileresults}, can be a significant source of support against the pull of gravity.  Because the velocity dispersion of the gas is relatively high ($\sim$100-150 km/s), it is capable of providing support especially in the outer region of the disk where the surface density profile steepens, enhancing the strength of the pressure gradient.  As shown in Figure~\ref{fig:tileresults}, we find that this additional source of pressure can reduce estimates of enclosed mass by $\sim$20\% or more in the outer disk.  While the turbulent pressure gradient is always important at large radii, we see mild evidence that at high redshift ($z > 2$) it can be of greater significance within the main body of the disk itself.  In the first two rows of Figure~\ref{fig:tileresults} ($z > 2$), the estimated mass from rotational velocity $M_{v_\phi}$ differs from $M_{\rm enc}$ by at least 10-20\% nearly everywhere, whereas in the lower two ($z < 2$) there is usually a region where the two match. 
The large effects of pressure gradients are especially important to model as accurately as possible since their radial dependence can change not only the normalization but also the \emph{shape} of outer rotation curves. In particular, pressure gradients tend to induce downturns in outer rotation curves which, if not modeled correctly, can spuriously imply a declining circular velocity.

We note that the gas velocity dispersions we measure here are higher than typically expected from observations \citep{Ubler2019}, although velocity dispersions of $\sim100$ km/s are not unheard of at this mass and redshift \citep{Price2019, Molina2019, Genzel2017}.  As a result, the magnitude of the effect of turbulent pressure gradients in these simulated galaxies may be somewhat elevated compared to real galaxies.  The gas density gradient in these systems, however, also plays a role in determining the net effect of ISM pressure.

In our analysis, we are able to measure the true physical force $1/\rho~dP/dr$ because we have access to the full three-dimensional information in the simulation.  Observationally, however, this force must be estimated using observable quantities such as line-of-sight velocity dispersion $\sigma_{\rm 1d}$ and gas surface density $\Sigma$.  

Surface density $\Sigma$, three-dimensional density $\rho$, and disk scale height $z_h$ are related as $\Sigma \sim \rho z_h$, such that their radial gradients are related as
\begin{equation}
\frac{d \log \Sigma}{d \log r} = \frac{d \log \rho}{d \log r} + \frac{d \log z_h}{d \log r}.
\end{equation}
A variety of assumptions about disk structure may be made in order to transform the gradient in $\rho$ to a gradient in $\Sigma$.  If scale height is fixed, $d 
\log z_h / d \log r = 0$ and the two gradients are identical.  Alternatively, under the assumptions that $\sigma$ is independent of height and the disk is in vertical hydrostatic equilibrium, the Spitzer solution for the vertical profile of an isothermal disk supplies a relationship between midplane density $\rho_0$ and surface density $\rho_0 \sim \Sigma^2$ such that $d \log \rho / d\log r = 2~d\log\Sigma/d\log r$.

In Figure \ref{fig:pressureestimators}, we test several variations of observational estimates of the turbulent pressure force of the form
\begin{equation}
\Delta v^2_{\rm press} = -\sigma^2 \frac{d \log (\rho \sigma^2)}{d\log r} \approx -2 \sigma^2 \frac{d \log \Sigma}{d \log r},
\end{equation}
a commonly-used assumption for massive high-redshift disks \citep[see e.g.][]{Burkert2016, Genzel2017},
against the true effect of turbulent pressure measured directly from the simulations.  In the bottom two rows, the blue lines (identical to the blue lines in Figure~\ref{fig:tileresults}) show the average effect on the estimated mass due to turbulent pressure $\Delta M_{\rm press}/M_{\rm enc}$ over a time period $\Delta z = 0.5$ for a given simulation, and thin black lines show observational estimates of the form above.  The top row shows the estimated enclosed mass relative to the true enclosed mass at that radius, where the purple lines are again identical to Figure~\ref{fig:tileresults}.  Here we assume for simplicity that the gravitational correction, $\Delta M_{\rm grav}$, has been correctly included.

The dot-dashed line uses $\sigma_{v_r}$ measured over an entire annulus and a double power law fit to the surface density profile for $d \log \Sigma / d \log r$.  As described elsewhere in this paper, we find that the double power law is generically a good fit to the profiles measured from the simulation, with inner slopes between 0 and -1 and outer slopes between -3 and -5.  This estimate systematically under-predicts the pressure support in the main body of the disk but typically matches or even exceeds the true pressure support in the outer disk.  The solid line uses the same definition of $\sigma$, but fits the surface density profile to an exponential form $\Sigma \propto e^{r/R_d}$.  Even though we have found that the exponential profile typically gives a poorer fit to the surface density profile than a double power law, the resulting estimate for the pressure support matches the true pressure support about as well (and sometimes even better) in the outer disk.
 Finally, the dashed line uses the exponential fit, but uses a measurement for $\sigma$ which is more local.  Rather than taking a measurement over the whole annulus, the annulus is segmented into 16 $\phi$-bins and the velocity dispersion is calculated within each bin relative to the mean radial velocity in that bin, rather than relative to the mean radial velocity of the annulus.  This method of measuring velocity dispersion will produce a lower value by definition, and we find that using this measurement therefore under-predicts the pressure support.  This exercise illustrates the danger of measuring velocity dispersion in too-small apertures which do not capture the largest scales of turbulent motion.

All three quasi-observational estimates systematically under-predict the turbulent pressure support within the main body of the disk.  The bottom row of panels illustrates the reason why.  Here, the effect measured from the simulation is overlaid with $-\sigma_{v_r}^2/v_c^2 \times \alpha$, where $\alpha = [1, 2, 3, 4]$ represents the logarithmic slope of the pressure profile $d \log P / d \log r$.  Here we see that the pressure profile in the main body of the disk typically has a slope between -1 and -2 (i.e., the blue lines lie between the black and purple lines), and steepens to slopes of -3 to -4 in the outer part of the disk.  Observationally, the slope of the pressure profile is usually estimated as some multiple of the surface density profile.  The surface density profiles, however, are very flat, with slopes close to zero in many cases (and indeed the exponential disk form assumes a flat profile in the center by definition).  Thus, the mild trend we see in the simulations for $\Delta M_{\rm press}$ becoming significant within the main body of the disk at higher $z$ would not be measurable using observational estimates which rely solely on surface density gradients.  The mismatch between these quasi-observational estimates and the directly-measured pressure gradient points to a breakdown in one or more of the assumptions made to transform $d\log P/ d \log r$ into $2 \times d \log \Sigma/ d \log r$, e.g. that $\sigma$ is independent of height or that the gas is in vertical hydrostatic equilibrium.

\subsection{Additional geometric effects}
\label{ssec:geometrydiscussion}

With the full simulation data, we are able to directly measure the gravitational potential felt by test particles in the disk, and compare it to the assumption of a spherically-symmetric potential. 
Unlike in observational studies, which typically assume a parametric form (e.g., a disk+bulge model),
our gravitational potential calculation is free of any assumptions about the way the mass is distributed. 
We find that the correction to the gravitational potential takes the same qualitative shape as for a \citet{Freeman1970} disk, reducing the gas rotational velocity in the central region and slightly enhancing it elsewhere relative to a spherical potential.

An additional important factor for dynamical mass inferences is how circular the orbits are, since a non-circular orbit means that the orbital velocity has components in both the $r$ and $\phi$ directions, and so $v_{\phi}$ will in general differ from $\sqrt{GM_{\rm enc}/r}$. In our simulated disks, the orbits of particles are circular in the main body of the disk but can be highly non-circular in the inner parts (due to a bar or other non-axisymmetric structure) and in the very outer region (where material is accreting from the cosmic web and the disk is vulnerable to external perturbations).  The accurate inference of enclosed mass depends upon the non-circular regions being identified and excluded.  In the simulations, we have the full 3D velocity data to do so, but observationally this is more challenging.  With full 2D kinematic maps, structures such as the quadrupole visible in Fig. 4 can be identified if the galaxy is inclined with respect to the observer (see \citet{Oman2017} for an example). 

\section{Summary and Conclusion}
\label{sec:conclusion}

In this study, we have measured the kinematics of gas in simulated disk galaxies and examined the physical effects that must be accounted for in order to infer the mass profile from those kinematic properties.  We have focused in particular on the case of massive high-redshift galaxies since these galaxies show elevated ISM turbulence relative to low-redshift disks.
Moreover, detailed, spatially-resolved kinematic information has only recently become available from large high-redshift samples.

Using four zoom-in simulations run with the FIRE-2 galaxy formation model which reach stellar masses of a few $\times 10^{11}$ \Msun~ at $z=1$ \citep{Angles-Alcazar2017}, we find that the following conditions must be met in order to accurately infer dynamical mass from gas rotation measurements:
\begin{itemize}
\item \textbf{A well-organized disk:} The gas must be in rotation-dominated motion, and have a smooth surface density profile without jumps or gaps.
\item \textbf{Circular orbits:} Any significant bulk motions in the $r$- or $z$- direction arising from inflows, outflows, or eccentric orbits in e.g. bars mean that $M$ and $v_\phi$ cannot be directly related to one another.
\item \textbf{Turbulent pressure taken into account:} The high gas velocity dispersions (100-150 km/s) in the simulated galaxies lead to strong turbulent pressures.  Radial gradients in turbulent pressure can provide significant support against central gravity, which reduces gas rotational velocity.
\item \textbf{Non-spherical mass distribution taken into account:} The disky component of the mass distribution modifies the gravitational potential such that $v_c \neq \sqrt{GM_{\rm enc}/r}$.
\end{itemize}

When the above conditions are satisfied ($\sim 40$\% of snapshots from $z=1-3$), we find that:
\begin{itemize}
\item There is an intermediate region (typically from $r \approx$ 1-3 kpc for the massive, high-redshift simulations but which may lie elsewhere for other systems) where the gas rotation profile matches the
spherically symmetric, zero-pressure expectation
$\bar{v}_\phi = \sqrt{GM_{\rm enc}/r}$.
\item Beyond the intermediate region (corresponding to a significant gradient in the gas surface density profile in the disk), $\bar{v}_\phi$ falls off relative to $\sqrt{GM_{\rm enc}/r}$ due to significant turbulent pressure support.
The effect on estimates of the enclosed mass $\Delta M_{\rm press}$ is radially-dependent. If not accounted for, this effect can significantly bias enclosed mass estimates low.
Because the effect is radially-dependent, it can change the {\it shape} of a gas rotation curve, in addition to its normalization.
\item At $z > 2$, $\Delta M_{\rm press}$ can be significant even within the main body of the disk and bias dynamical mass estimates low throughout.
\item Observational estimates of the pressure support can capture the magnitude of the pressure support in the outer disk, but may underestimate its effect at smaller radii.
\item The correction due to asphericity of the gravitational potential $\Delta M_{\rm grav}$ is most important in the center of the disk and can reduce the estimated mass.
\item On average, the total mass profile may be successfully recovered when all these effects are accounted for.
\end{itemize}

Our current simulations do not include a model for supermassive black hole feedback. This likely leads quantitatively to higher central densities and rotational velocities than expected, on average, for observed galaxies in the high-mass regime where AGN can have dynamically important effects.  However, in the Appendix 
we show that our physical analysis and qualitative results also hold for simulations of lower-mass and lower-redshift galaxies whose structural and kinematic properties are in better agreement with observations.

In general, we find that when turbulent pressure gradients, non-spherical potentials, and non-circular orbital structures are appropriately taken into account, measures of rotation in high-z massive galaxies may be used to recover an estimate of the enclosed mass.  This implies that one may be able to accurately recover enclosed mass from quantities available in observational data.  Some of the physical effects we discuss, however, may be challenging to account for at high redshift, where the spatial resolution of the measurements is limited (especially the identification of inner regions where the orbits can be highly non-circular).  We also note that we have not attempted in this paper to account for other, purely observational effects such as low resolution, beam smearing, and inclination corrections. Modeling these observational effects on dynamical mass measurements 
will be important to fully realize the potential of recent and upcoming instruments capable of spatially resolving galaxy kinematics at high redshift.

\section*{Data Availability}

The data supporting the plots within this article are available on reasonable request to the corresponding author. A public version of the GIZMO code is available at http://www.tapir.caltech.edu/~phopkins/Site/GIZMO.html. 
Additional data including simulation snapshots, initial conditions, and derived data products are available at https://fire.northwestern.edu/data/.

\section*{Acknowledgements}

SW is supported as a CIERA Fellow by the CIERA Postdoctoral Fellowship Program (Center for Interdisciplinary Exploration and Research in Astrophysics, Northwestern University).
CAFG was supported by NSF through grants AST-1517491, AST-1715216, and CAREER award AST-1652522; by NASA through grant 17-ATP17-0067; by STScI through grant HST-AR-14562.001; and by a Cottrell Scholar Award from the Research Corporation for Science Advancement.
DAA acknowledges support by the Flatiron Institute, which is supported by the Simons Foundation.
RF acknowledges financial support from the Swiss National Science Foundation (grant no 157591).
Support for PFH was provided by an Alfred P. Sloan Research
Fellowship, NSF Collaborative Research Grant \#1715847, and
CAREER grant \#1455342, and NASA grants NNX15AT06G, JPL
1589742, 17-ATP17-0214.
DK was supported by NSF grant AST-1715101 and the Cottrell Scholar Award from the Research Corporation for Science Advancement.
Numerical calculations were run on the Quest computing cluster at Northwestern University; the Wheeler computing cluster at Caltech; XSEDE allocations TG-AST130039, TG-AST120025, TG-AST140023, and TG-AST160048; and Blue Waters PRAC allocation NSF.1713353.

\bibliography{biblio.bib}

\appendix

\section{Analysis for z=0 Milky-Way-mass galaxies}
\label{app:m12s}

\begin{figure*}
  \centering
  \includegraphics[width=1.5\columnwidth]{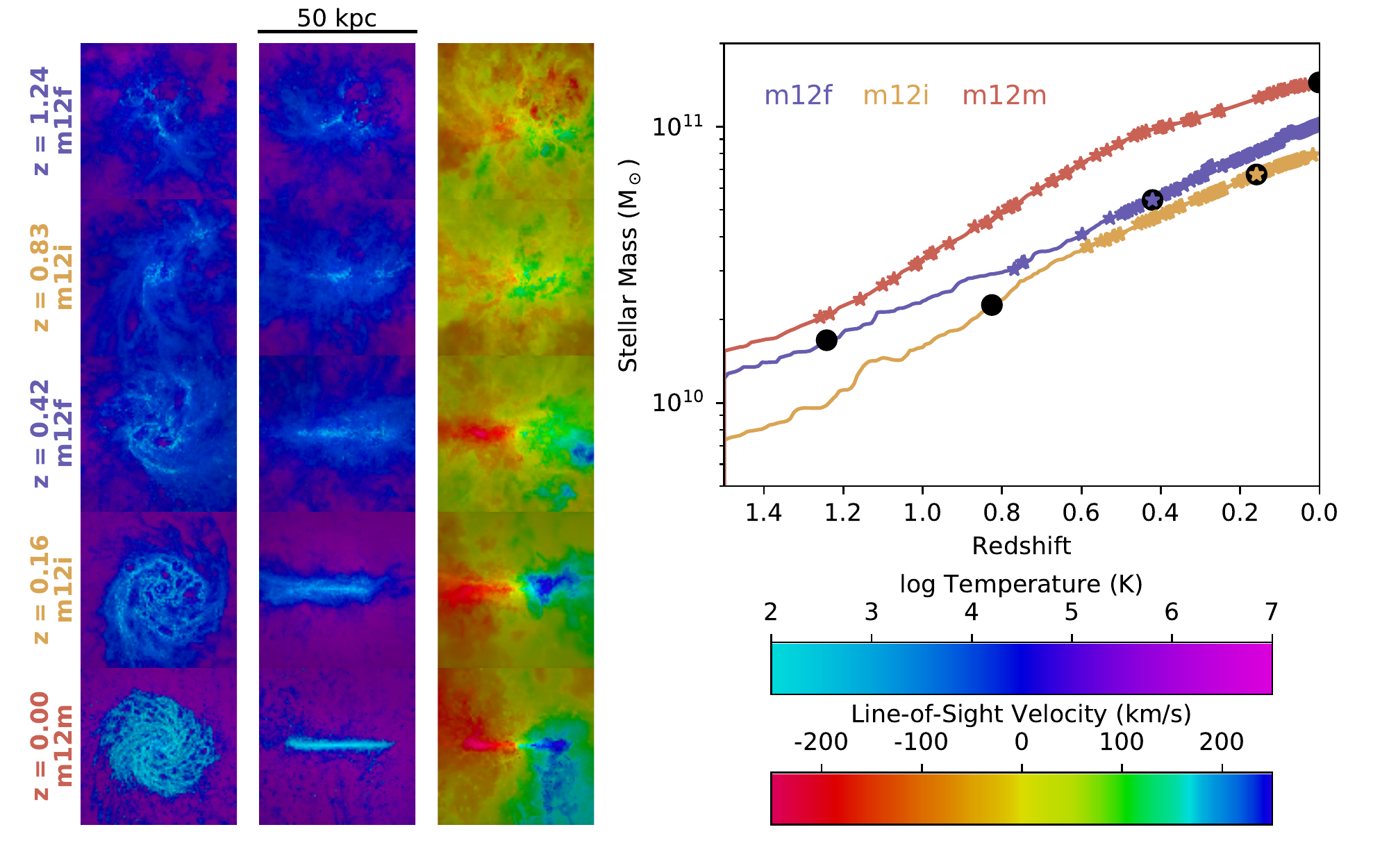}
  \caption{Mass evolution and snapshot images of three simulated Milky-Way-mass disks, analogous to Figure~\ref{fig:massevol}.  \textbf{Right:} Stellar mass evolution from $z=1.5$ to $z=0$.  Star symbols indicate the presence of a well-behaved gaseous disk (as defined in Section~\ref{ssec:calcs}) at that snapshot.  Black circles mark the snapshots whose images are shown on the left.  \textbf{Left:}  Images of average gas temperature (left and center columns) and line-of-sight velocity (right column) for the snapshots marked with black circles.  The leftmost column shows a face-on view while the center and right columns are edge-on.}
  \label{fig:m12intro}
\end{figure*}

The results presented in the main body of the text pertain primarily to high-redshift massive galaxies, in which the gas kinematics can be particularly strongly affected by elevated ISM turbulence. 
In this Appendix, we repeat our analysis on a set of simulations of lower-redshift and lower-mass systems: Milky-Way-like galaxies down to $z=0$.  The properties of these simulations (known by the names m12f, m12i, and m12m) have been analyzed in previous papers \citep[e.g.,][]{Wetzel2016, Garrison-Kimmel2017, Sanderson2018, El-Badry2018}. 
These simulations were run using the same FIRE-2 hydrodynamical code and galaxy formation model described in Section~\ref{ssec:sims}, with the exception that they do not include supermassive black hole formation and accretion (which are not important for our analysis).  The mass resolution of these simulations is $m_{\rm b}=7 \times 10^3$ \Msun~for baryons and $m_{\rm DM}=3.5 \times 10^4$ \Msun~for dark matter.    

\begin{figure}
  \centering
  \includegraphics[width=\columnwidth]{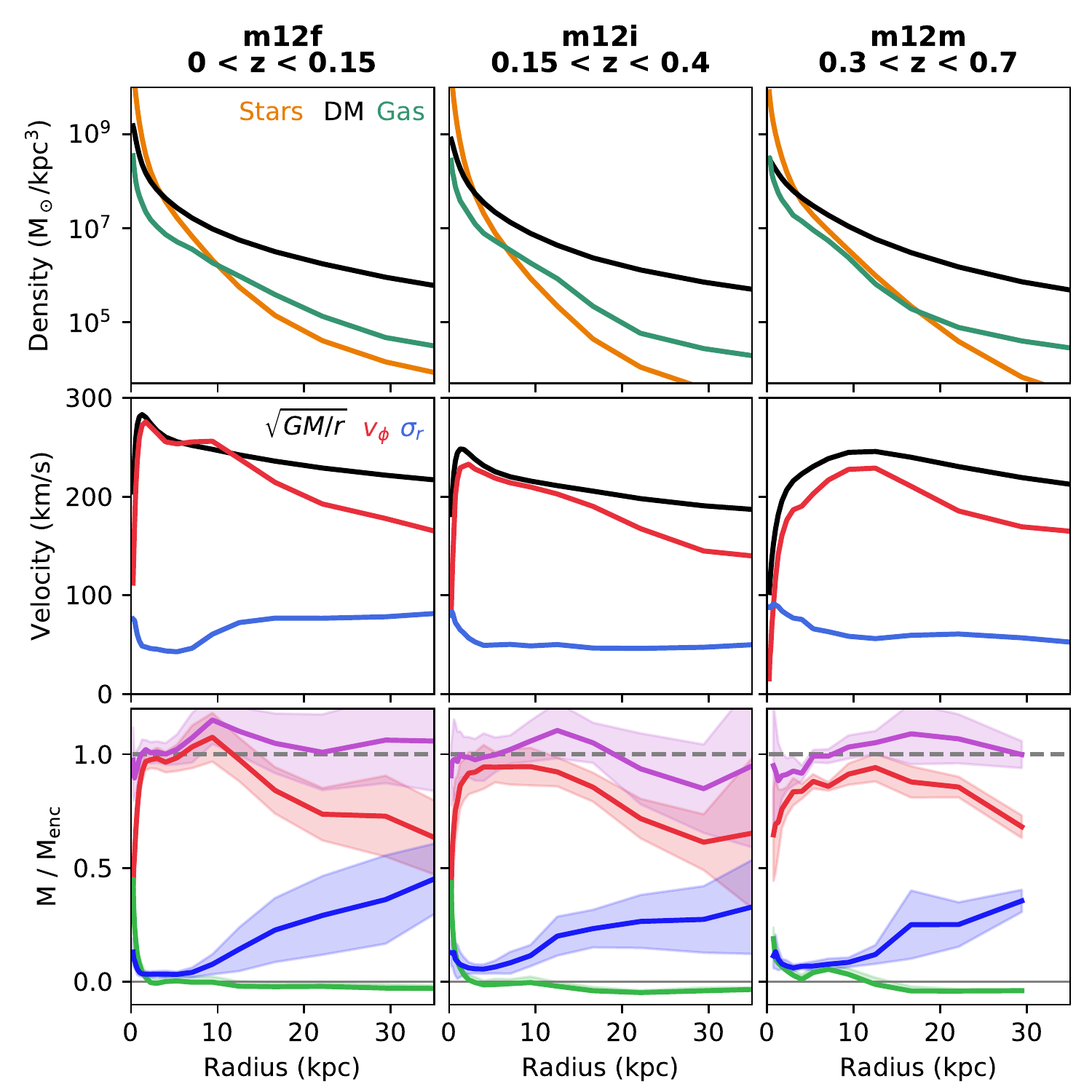}
  \caption{Profiles of density (top row), velocity (middle row), and estimated mass (bottom row) for the simulated Milky-Way-mass halos at a variety of redshifts, analogous to Figures~\ref{fig:allprofs} and \ref{fig:tileresults}.  Each panel shows an average over the redshift range indicated at the top of the column.}
  \label{fig:m12profs}
\end{figure}

The stellar mass evolution of m12f, m12i, and m12m appears in the right side of Figure~\ref{fig:m12intro}.  As in Figure~\ref{fig:massevol}, snapshots where the gaseous disk is ``well-behaved" according to the criteria laid out in Section~\ref{ssec:calcs} are marked with a star symbol.   These lower-mass, later-forming galaxies do not form persistent disks until $z \lesssim 0.5$, although m12m does sporadically form a disk at earlier times.  The final stellar masses of these three simulations range from $0.6 - 1.5 \times 10^{11}$ \Msun.  Images of individual snapshots (marked with black circles in the mass evolution panel) are shown on the left side of Figure~\ref{fig:m12intro}.  The first two columns display the gas temperature from face-on and edge-on viewpoints according to the angular momentum axis of the gas, and the third column replicates the edge-on view with color instead indicating line-of-sight velocity.  These snapshots demonstrate the systems' general evolution from messy, bursty star formation to well-organized disks.

Density and velocity profiles for the three simulations over different redshift ranges appear in the top two rows of Figure~\ref{fig:m12profs}.  These profiles broadly mirror the patterns seen in the main body of the text (though with lower densities and velocities overall), with the stellar component dominant in the central few kpc and a transition to dark matter dominance further from the galaxy center.  These disks are generally more extended than the massive disks discussed in the main body of the text.  The velocity profiles of these galaxies peak in the 200-300 km/s range, and are flatter than the velocity profiles of the high-redshift massive galaxies because their baryonic mass is less concentrated at the center.  We measure a gas velocity dispersion of $\sim$40 km/s in the main body of the disk which increases toward the galaxy center.  This velocity dispersion is somewhat higher than might be expected from observations of Milky-Way-mass galaxies in the local Universe, but is consistent with the measurements made from the same simulations by \citet{El-Badry2018}.

As seen in the high-redshift massive galaxies, the rotational velocity of the gas $\bar{v}_\phi$ is well-matched to $v_c = \sqrt{GM_{\rm enc}/r}$ in the main body of the disk, but falls off at inner and outer radii.  This well-matched region, however, is more extensive than for the galaxies studied in the main text, covering a region from $r \approx$ 1-10 kpc.  This is consistent with our observational expectation that radial gradients in turbulent pressure should not play a large role in supporting Milky-Way-mass galaxies today.  

In the bottom row of Figure~\ref{fig:m12profs}, we repeat our analysis measuring the effect of turbulent pressure support and non-spherical gravitational potentials on estimates of dynamical mass.  Here, we demonstrate that our methodology is also applicable and reliable for galaxies in a completely different mass and redshift regime.  The results are qualitatively similar to those seen for the massive galaxies: pressure gradients become significant in the outer disk, the effect of non-sphericity is strongest in the center, and accounting for these effects (in addition to requiring disk smoothness and orbital circularity) enables the recovery of the true dynamical mass across radii.  We also observe the same mild redshift trend seen in the high-redshift massive galaxies where pressure support becomes increasingly important in the main body of the disk at higher $z$.

\label{lastpage}

\end{document}